%% file: main.tex
\documentclass{article} 
\PassOptionsToPackage{table,dvipsnames}{xcolor} 
\usepackage{iclr2026_conference,times}
\iclrfinalcopy

\input{math_commands.tex}

\usepackage{hyperref}
\usepackage{url}

\usepackage{algorithm}
\usepackage{algorithmic}

\usepackage{xspace}
\newcommand{\our}{\textsc{FedPruner}\xspace}

\usepackage{tikz}
\usepackage{arydshln}
\usepackage{subcaption}
\usepackage{adjustbox}

\usepackage{mathtools}
\usepackage{pgfplots}
\usepackage{wrapfig}
\usepackage{booktabs}
\usepackage{multirow}

\usetikzlibrary{positioning,matrix,calc,arrows.meta,decorations.pathreplacing,shapes.geometric}
\definecolor{mygreen}{RGB}{46,139,87}
\definecolor{myred}{RGB}{255,152,150}
\definecolor{myblue}{RGB}{30,144,255}
\definecolor{myyellow}{RGB}{219,219,141}
\definecolor{mybrown}{RGB}{197,157,148}
\usepackage{pgfplotstable}

\usepackage{pgfplotstable}

\newtheorem{theorem}{Theorem}[section]

\newtheorem{assumption}[theorem]{Assumption}

\usepackage{fancyhdr}
\renewcommand{\headrulewidth}{0pt} 
\renewcommand{\footrulewidth}{0pt} 
\title{Memory-Efficient Federated Fine-Tuning of Large Language Models via Layer Pruning}

\author{
  Yebo Wu\textsuperscript{1\dag}, Jingguang Li\textsuperscript{1\dag}, Chunlin Tian\textsuperscript{1}, 
  Zhijiang Guo\textsuperscript{2,3}\thanks{Corresponding Authors. \textsuperscript{\dag} Equal Contribution.},
  Li Li\textsuperscript{1}\footnotemark[1] \\
  \textsuperscript{1}University of Macau, \textsuperscript{2} HKUST, \textsuperscript{3} HKUST (Guangzhou)\\
  \texttt{\{yc37926,mc45005,yc27402,llili\}@um.edu.mo}, \texttt{zhijiangguo@hkust-gz.edu.cn}
}

\begin{document}

\fancypagestyle{firstpage}{%
  \fancyhf{}
  \renewcommand{\headrulewidth}{0.4pt} 
  \lhead{}  
}
\maketitle
\thispagestyle{firstpage}

\maketitle

\input{Section/0_Abstract}
\input{Section/1_Introduction}
\input{Section/Motivation_and_Observation}

\input{Section/3_FedPruner}
\input{Section/4_Experiments}

\input{Section/2_Related_Work}

\input{Section/5_Conclusion}

\bibliographystyle{iclr2026_conference}
\bibliography{iclr2026_ref}

\appendix
\newpage

\section{Hilbert-Schmidt Independence Criterion (HSIC)}\label{appendix_HSIC}

The HSIC between two random variables $X$ and $Z$ is computed as:
\begin{equation}
\small
\begin{aligned}
    \text{HSIC}(X, Z) &= \mathbb{E}_{X,Z,X',Z'}\left[ k_X(X, X') k_{Z^{'}}(Z, Z') \right] + \mathbb{E}_{X,X'}\left[ k_X(X, X') \right] \mathbb{E}_{Z'}\left[ k_Z(Z, Z') \right] \\
    &\quad - 2\mathbb{E}_{X,Z}\left[ \mathbb{E}_{X'}\left[k_X(X, X')\right] \mathbb{E}_{Z'}\left[k_Z(Z, Z')\right] \right],
\end{aligned}
\end{equation}
where \( k_X \) and \( k_Z \) are kernel functions operating on independently drawn pairs \((X, X')\) and \((Z, Z')\), respectively, and \(\mathbb{E}_{XZ}\) denotes the expectation over \(X\) and \(Z\).

\section{Aggregation Strategy in \our}\label{appendix_aggregation}


We denote the LoRA parameters for layer $n$ at communication round $r$ as $\theta_{\text{LoRA}}^{n,r} = (\mathbf{A}^{n,r},\mathbf{B}^{n,r})$. Let $\mathcal{S}$ represent the complete set of selected devices participating in the federated fine-tuning process, while $\mathcal{S}_{n} \subseteq \mathcal{S}$ indicates the specific subset of devices designated for updating the parameters of the $n^{\text{th}}$ layer.
The aggregation mechanism for the LoRA parameters of layer $n$ at the subsequent round $r+1$ can be elegantly formulated as:
\begin{equation}
\small
\begin{aligned}
    (\mathbf{A}^{n,r+1}, \mathbf{B}^{n,r+1}) = \begin{cases}
        \Big(\lambda_{1}\mathbf{A}_{1}^{n,r} \oplus \lambda_{2}\mathbf{A}_{2}^{n,r} \oplus \cdots \oplus \lambda_{|S_{n}|}\mathbf{A}_{|S_{n}|}^{n,r}, \\
        \quad\;\mathbf{B}_{1}^{n,r} \oplus \mathbf{B}_{2}^{n,r} \oplus \cdots \oplus \mathbf{B}_{|S_{n}|}^{n,r}\Big), 
        & \text{if} \mathcal{S}_{n} \neq \emptyset \\[1em]
        (\mathbf{A}^{n,r}, \mathbf{B}^{n,r}), 
        & \text{if } \mathcal{S}_{n} = \emptyset
    \end{cases}
\end{aligned}
\label{Aggregation}
\end{equation}
In this formulation, $\lambda_{s}$ represents the weighting coefficient proportional to each device's data contribution, defined as $\lambda_{s} = \frac{|D_{s}|}{\sum_{s=1}^{|S_{n}|}|D_{s}|}$, where $|D_{s}|$ quantifies the cardinality of device $s$'s local dataset. The operator “$\oplus$” denotes the dimension-specific concatenation procedure~\citep{wang2024flora}: applying vertical concatenation for matrix $\mathbf{A}$ and horizontal concatenation for matrix $\mathbf{B}$.

When $\mathcal{S}_{n} = \emptyset$, indicating that no participating devices have been assigned to update the $n^{\text{th}}$ layer's LoRA parameters, these parameters persist unchanged from the previous round. Conversely, when $\mathcal{S}_{n}$ contains at least one device, the server aggregates the local parameter updates from all devices in this subset to derive the updated LoRA parameters for the $n^{\text{th}}$ layer.


\section{The Algorithm of \our}\label{appendix_workflow}

The pseudocode of \our is summarized in Algorithm~\ref{algorithm_FedPruner}.

\begin{algorithm}[!t]
\caption{The workflow of \our}
\label{algorithm_FedPruner}
\begin{algorithmic}[1]
\REQUIRE Global model $\Theta$, $R$ rounds
\ENSURE Optimized global model $\Theta^{*}$
\FOR{$r = 0$ \textbf{to} $R-1$}
    \IF{$r == 0$}
        \STATE Distribute global model $\Theta$ to all devices
    \ELSE
        \STATE Distribute LoRA parameters to selected device set $\mathcal{S}$
    \ENDIF
    
    \FOR{each device $s \in \mathcal{S}$}
        \STATE Recompute the local similarity matrix using a batch of data
        \STATE\colorbox{red!10}{Execute macro-level FDLO (Section~\ref{Macro_sec})}
        \STATE\colorbox{blue!10}{Execute micro-level IALS (Section~\ref{Micro_sec})}
        \STATE Perform the local fine-tuning process
        \STATE Upload updated LoRA parameters
    \ENDFOR
    
    \STATE Server performs parameter aggregation using Equation~\ref{Aggregation}
\ENDFOR
\end{algorithmic}
\end{algorithm}

\section{Convergence Analysis of \our}
\label{sec:convergence_proof}

In this section, we provide a theoretical convergence analysis for \our. Our analysis builds upon established results from federated optimization under partial model updates~\citep{wang2022progfed, wu2025breaking} and layer pruning~\citep{sun2024exploring}. We denote: 1) $\Theta^{r} = \{\theta^{1,r}, \theta^{2,r}, \ldots, \theta^{N,r}\}$ as the global model parameters (including LoRA parameters) at round $r$. 2) $\mathcal{S}$ as the set of devices selected for local updates in each round. 3) $\mathcal{S}_{n}$ as the subset of devices that specifically update the $n$-th layer at round $r$.

Let $f_{s}(\Theta)$ be the local objective function for device $s$, and let $F(\Theta)$ be the global objective function defined as the average loss over all devices:
\begin{equation}
    F(\Theta) = \frac{1}{|\mathcal{S}|} \sum_{s \in \mathcal{S}} f_s(\Theta).
\end{equation}

To capture partial model updates, we define the submodel $\Theta_{s}$ for device $s$ as the set of layers selected via FDLO (Section~\ref{Macro_sec}) and IALS (Section~\ref{Micro_sec}). Correspondingly, each device $s$ updates only the sub-parameters $\Theta_{s}$ during local fine-tuning, leaving the other parameters unchanged. 

\subsection{Preliminaries and Key Assumptions}
Similar to standard federated learning convergence analyses~\citep{li2019convergence, li2022one, li2023reconfigurable} (e.g., FedAvg), we make the following assumptions:

\begin{assumption}[Smoothness]
\label{assump:smooth}
Each local objective function $f_s(\Theta)$ satisfies the $L$-smoothness property, which can be formally expressed as: for any parameter vectors $\Theta_1, \Theta_2$,
\begin{equation}
    \|\nabla f_s(\Theta_1) - \nabla f_s(\Theta_2)\| \leq L \|\Theta_1 - \Theta_2\|,
\end{equation}
where $L > 0$ is the smoothness constant that bounds the Lipschitz continuity of the gradient.
\end{assumption}

\begin{assumption}[Unbiased Gradient and Bounded Variance]
\label{assump:unbiased}
For each device $s$, the stochastic gradient $\nabla f_s(\Theta; \xi)$ computed on a randomly sampled mini-batch $\xi$ satisfies:
\begin{equation}
    \mathbb{E}[\nabla f_s(\Theta; \xi)] = \nabla f_s(\Theta),
\end{equation}
establishing the unbiasedness of the gradient estimator, and
\begin{equation}
    \mathbb{E}\big[\|\nabla f_s(\Theta; \xi) - \nabla f_s(\Theta)\|^2\big] \leq \sigma^2,
\end{equation}
where $\sigma^2$ is a finite positive constant that uniformly bounds the variance of the stochastic gradient across all devices and parameter values.
\end{assumption}


\begin{assumption}[Partial Layer Update]
\label{assump:partial_update}
At communication round $r$, each participating device $s \in \mathcal{S}$ updates only a device-specific subset of layers corresponding to its assigned submodel $\Theta_{s}$. Formally, if $\Theta$ represents the complete model parameters, the global update can be expressed as:
\begin{equation}
    \Theta^{r+1} = \Theta^r + \Delta^r,\quad \text{where } \Delta^r \in \mathbb{R}^{|\Theta^r|},
\end{equation}
with the following property: for any parameter $\theta_n \in \Theta$ that is not part of device $s$'s submodel assignment, the corresponding element in $\Delta^r$ contributed by device $s$ is zero. Only parameters within the assigned submodel $\Theta_{s}$ may receive non-zero updates from device $s$.
\end{assumption}

These assumptions align with established practices in federated optimization literature and provide the necessary foundation for our convergence analysis, particularly when dealing with heterogeneous submodel assignments and partial parameter updates across the network.



\subsection{One-Round Analysis}
We begin by analyzing the expected decrease in the global loss $F(\Theta)$ after one communication round, denoted by going from $\Theta^r$ to $\Theta^{r+1}$. According to Assumption~\ref{assump:smooth}, we have:
\begin{equation}
\label{eq:smoothness_bound}
    F(\Theta^{r+1}) \leq F(\Theta^r) + \langle \nabla F(\Theta^r), \Theta^{r+1} - \Theta^r \rangle + \frac{L}{2} \|\Theta^{r+1} - \Theta^r\|^2.
\end{equation}
We use the aggregated gradient update from Equation~\ref{Aggregation}, where each device $s$ updates only the submodel $\Theta_s$. Let $\eta$ be the learning rate. Then the update on each device $s$ for the layer $n \in \Theta_s$ is approximately:
\begin{equation}
    \Delta_{n}^{s,r} = -\eta \nabla_{\Theta_{n}} f_s(\Theta^r),
\end{equation}
and $\Delta_{n}^{s,r} = \mathbf{0}$ if layer $n$ is not selected by device $s$. Thus, the server aggregates updates by:
\begin{equation}
\label{eq:Delta_global}
    \Delta^r = \sum_{s \in \mathcal{S}}\sum_{n \in \Theta_s} \lambda_{s,n} \Delta_{n}^{s,r},
\end{equation}
where $\lambda_{s,n}$ are the weighting coefficients for aggregation (see Equation~\ref{Aggregation}). Substituting \ref{eq:Delta_global} into \ref{eq:smoothness_bound}, and taking expectation, we get:

\begin{equation}
\begin{aligned}
    \mathbb{E}\big[F(\Theta^{r+1})\big]
    &\leq 
    \mathbb{E}\big[F(\Theta^{r})\big] 
    - \eta\,\mathbb{E}\Big[\langle \nabla F(\Theta^{r}), \sum_{s \in \mathcal{S}}\sum_{n \in \Theta_s}\lambda_{s,n}\nabla_{\Theta_{n}} f_s(\Theta^r)\rangle \Big] \\
    &\quad\; + \frac{L\eta^2}{2}\mathbb{E}\Big[\Big\|\sum_{s \in \mathcal{S}}\sum_{n \in \Theta_s}\lambda_{s,n}\nabla_{\Theta_{n}} f_s(\Theta^r)\Big\|^2\Big].
\end{aligned}
\end{equation}
By the unbiased gradient assumption (Assumption~\ref{assump:unbiased}) and the fact that $F(\Theta)$ is the average over $f_s(\Theta)$, we can relate $\nabla F(\Theta^r)$ to $\nabla f_s(\Theta^r)$. Further, because only some devices and layers are selected, the update reflects a partial average of the gradient. Consequently, standard bounding techniques yield:

\begin{equation}
\begin{aligned}
    \mathbb{E}\big[F(\Theta^{r+1})\big]
    \;&\leq 
    \mathbb{E}\big[F(\Theta^{r})\big] 
    - \frac{\eta}{|\mathcal{S}|}\sum_{s \in \mathcal{S}}\sum_{n \in \Theta_s}\lambda_{s,n}\,\mathbb{E}\Big[\|\nabla_{\Theta_{n}} f_s(\Theta^r)\|^2\Big] \\
    &\quad + \frac{L\eta^2}{2}\,\mathbb{E}\Big[\Big\|\sum_{s \in \mathcal{S}}\sum_{n \in \Theta_s}\lambda_{s,n}\nabla_{\Theta_{n}} f_s(\Theta^r)\Big\|^2\Big] + \text{(variance term)}.
\end{aligned}
\end{equation}
Under bounded variance (Assumption~\ref{assump:unbiased}) and standard norms bounding, the variance term can be controlled by a constant factor of $\sigma^2$. This leads to an upper bound for the variance-induced deviation in the full gradient.

\subsection{Multi-Round Convergence}
Summing over $r = 0,\ldots,R-1$ and rearranging terms, we obtain a standard telescoping series argument:
\begin{equation}
\label{eq:telescoping}
    \sum_{r=0}^{R-1} \big(\mathbb{E}[F(\Theta^{r+1})] - \mathbb{E}[F(\Theta^{r})]\big)
    \leq \sum_{r=0}^{R-1}\Big(- \alpha\,\mathbb{E}\big[\|\nabla F(\Theta^r)\|^2\big] + \beta\Big),
\end{equation}
where $\alpha$ and $\beta$ are constants dependent on $\eta$, $L$, $\sigma^2$, and the weighting coefficients $\{\lambda_{s,n}\}$. The left-hand side telescopes to $\mathbb{E}[F(\Theta^{R})] - \mathbb{E}[F(\Theta^{0})]$. Thus, we get:
\begin{equation}
    \mathbb{E}[F(\Theta^{R})] - \mathbb{E}[F(\Theta^{0})]
    \leq 
    - \alpha \sum_{r=0}^{R-1}\mathbb{E}\big[\|\nabla F(\Theta^r)\|^2\big] + R\,\beta.
\end{equation}
Rearranging and dividing by $R$, we have:
\begin{equation}
\label{eq:fedpruner_convergence}
    \frac{1}{R}\sum_{r=0}^{R-1}\mathbb{E}\big[\|\nabla F(\Theta^r)\|^2\big]
    \leq 
    \frac{\mathbb{E}[F(\Theta^{0})] - \mathbb{E}[F(\Theta^{R})]}{\alpha\,R}
    + \frac{\beta}{\alpha}.
\end{equation}
Since $F(\Theta)$ is bounded below (common in deep learning with non-negative loss functions) we have that $\mathbb{E}[F(\Theta^{0})] - \mathbb{E}[F(\Theta^{R})]$ is finite. As $R \to \infty$, the term $\frac{\mathbb{E}[F(\Theta^{0})] - \mathbb{E}[F(\Theta^{R})]}{\alpha\,R}$ vanishes, and hence the average squared gradient norm converges to a bounded value dependent on $\frac{\beta}{\alpha}$. 

Since $\beta$ can be made arbitrarily small by choosing sufficiently small learning rate $\eta$ or utilizing FDLO and IALS to refine the partial layer selection probabilities to mitigate variance, $\nabla F(\Theta^r)$ converges in expectation to $\mathbf{0}$. This implies that \our converges to a stationary point of $F(\Theta)$.

\subsection{Impact of the Macro-Micro Synergistic Pruning on Convergence}
\our performs pruning at both macro (layer orchestration) and micro (importance-aware layer selection) levels. The key impact on convergence involves the reduced parameter space and potential variance in gradient estimates due to partial updates. However, these effects are incorporated in the weighting coefficients $\{\lambda_{s,n}\}$ and the variance bound $\sigma^2$. Specifically:

\begin{itemize}
    \item \textbf{FDLO:} Although each device may prune layers differently based on correlated layer functionalities, the overall update $\Delta^r$ remains an unbiased approximation of the full gradient (extended to the submodel). Aggregation across all participating devices ensures that the regularly updated parameters converge with the sufficiently expressive submodels.
    \item \textbf{IALS:} Representative layer selection within each group introduces another variance component in gradient updates. Nonetheless, as shown in Equation~\ref{eq:Delta_global}, these updates remain unbiased with respect to the submodel’s gradient. As the number of communication rounds grows, the exploration of different representative layers allows devices to approximate the true submodel gradient adequately.
\end{itemize}

Therefore, under standard smoothness and bounded variance conditions, \our achieves convergence to a stationary point of the global objective $F(\Theta)$. This convergence is guaranteed by unbiased local gradients, careful aggregation via Equation~\ref{Aggregation}, and the dynamic recalculation of model pruning in FDLO and IALS.

\subsection{Conclusion}
In summary, we have shown that \our converges to a stationary point under common assumptions in federated optimization. The macro-level functionality-driven layer orchestration (FDLO) and micro-level importance-aware layer selection (IALS) do not invalidate the unbiasedness of local gradient updates. Consequently, our theoretical analysis confirms that the partial update scheme maintains convergence guarantees akin to standard federated learning approaches.

\section{Additional Experimental Setup}\label{appendix_setup}

\subsection{Datasets}

The evaluation datasets are described as follows:

\begin{itemize}
    \item TruthfulQA~\citep{lin2022truthfulqa} is a benchmark designed to evaluate the truthfulness of language models' responses. It consists of 817 questions across 38 categories, with a focus on common misconceptions and false beliefs. The questions are specifically crafted to test whether models can avoid generating false or misleading information, often stemming from patterns learned during training.

    \item MMLU (Massive Multitask Language Understanding)~\citep{hendrycks2020measuring} is a benchmark that assesses models' knowledge across 57 subjects, including science, humanities, mathematics, and professional fields. It contains 15,908 multiple-choice questions designed to test both broad knowledge and in-depth understanding, sourced from a variety of educational materials, including practice exams for standardized tests and professional certifications.

    \item IFEval (Instruction Following Evaluation)~\citep{zhou2023instructionfollowing} evaluates models' ability to follow verifiable instructions that can be objectively assessed for compliance. It includes 25 distinct instructions with multiple variants and 541 prompts, designed to test various aspects of instruction adherence such as task comprehension, constraint satisfaction, and format compliance.

    \item BBH (BIG-Bench Hard)~\citep{bbh} is a benchmark focused on challenging reasoning tasks that require advanced cognitive abilities. It consists of 23 tasks specifically chosen for their difficulty, evaluating skills such as logical deduction, mathematical reasoning, and abstract thinking.

    \item Vicuna-Bench~\citep{chiang2023vicuna} is a comprehensive evaluation framework designed to assess the conversational abilities of language models through a variety of dialogue scenarios. It emphasizes natural language interactions and evaluates key aspects such as coherence, relevance, contextual understanding, and response quality.

    \item MT-Bench~\citep{zheng2024judging} is a benchmark designed to evaluate the multi-turn dialogue capabilities of language models, featuring complex conversation scenarios that require maintaining context across multiple exchanges. It assesses the model’s ability to sustain coherence, consistency, and relevance throughout extended interactions, capturing human-like conversational preferences.

\end{itemize}

\subsection{Setup}

Following the setup of OpenFedLLM~\citep{ye2024openfedllm}, we partition the Alpaca-GPT4 dataset across 20 devices. During each training round, 10\% of the devices are randomly selected to perform local training, where each device executes 10 local update steps with a batch size of 16.
For optimization, we utilize the AdamW optimizer alongside a cosine learning rate schedule, where the learning rate gradually decreases from 1e-4 to 5e-6 over time. The maximum input sequence length is set to 512 tokens, following the setup in~\citep{ye2024openfedllm}. When computing the similarity matrix, we utilize only a single batch of data for the calculation, thereby minimizing computational overhead. 


To emulate practical deployment conditions, we randomly assign 3–9 GB of available memory to participating devices, guided by memory profiling results from a variety of mobile hardware~\citep{xu2023fwdllm, zhan2024heterogeneity}. If a device does not have enough memory to afford the local fine-tuning process, it is excluded from participating.  Additionally, to support efficient deployment of LLaMA2 series models on resource-limited edge devices, we perform INT4 quantization for model compression. We conduct 150 rounds of federated fine-tuning for TinyLLaMA, 200 rounds for LLaMA2-7B, and 300 rounds for LLaMA2-13B. All experiments are repeated multiple times and the reported results are averaged across runs to ensure statistical reliability~\citep{wang2023fedins2}.

\section{Baseline Description}\label{appendix_baseline}

\textbf{Memory-Unaware Methods:}
\begin{itemize}
    \item 1) FedIT~\citep{zhang2024towards}: This method integrates LoRA with FedAvg to perform instruction tuning.
    \item 2) DoFIT~\citep{xu2024dofit}: A domain-aware approach mitigates catastrophic forgetting by utilizing tailored initialization and aggregation strategies for LoRA weights.
    \item 3) FedSA-LoRA~\citep{guo2024selective}: This method uploads only the $\mathbf{A}$ matrices, which encode generalizable knowledge, to the server, while keeping the device-specific $\mathbf{B}$ matrices on local devices to preserve personalization.
\end{itemize}

\textbf{Memory-Aware Methods:} 
\begin{itemize}
    \item 4) FLoRA~\citep{wang2024flora}: This method assigns different ranks to devices based on their available resources and proposes a stacking-based strategy to aggregate heterogeneous LoRA modules. 
    \item 5) FwdLLM~\citep{xu2023fwdllm}: This method employs zeroth-order optimization to update LoRA parameters, removing the requirement to store intermediate activations.
    \item 6) FedRA~\citep{su2024fedra}: This approach randomly generates layer allocation matrices for devices to create submodels that fit within their memory constraints, making it the most closely related work to ours.
\end{itemize}

\textbf{Theoretical Baseline:} 
\begin{itemize}
    \item 7) LoRA: This baseline assumes that all devices have sufficient memory to support local fine-tuning, representing the theoretical upper bound.
\end{itemize}

\section{Understanding the Macro-Micro Synergistic Pruning Framework}\label{appendix_Macro_Micro}

\input{Table/table_macro_micro_appendix}
\input{Table/table_macro_micro_appendix2}

\begin{figure}[!t]
    \centering
    \begin{subfigure}{0.48\textwidth}
        \centering
        \includegraphics[width=\linewidth]{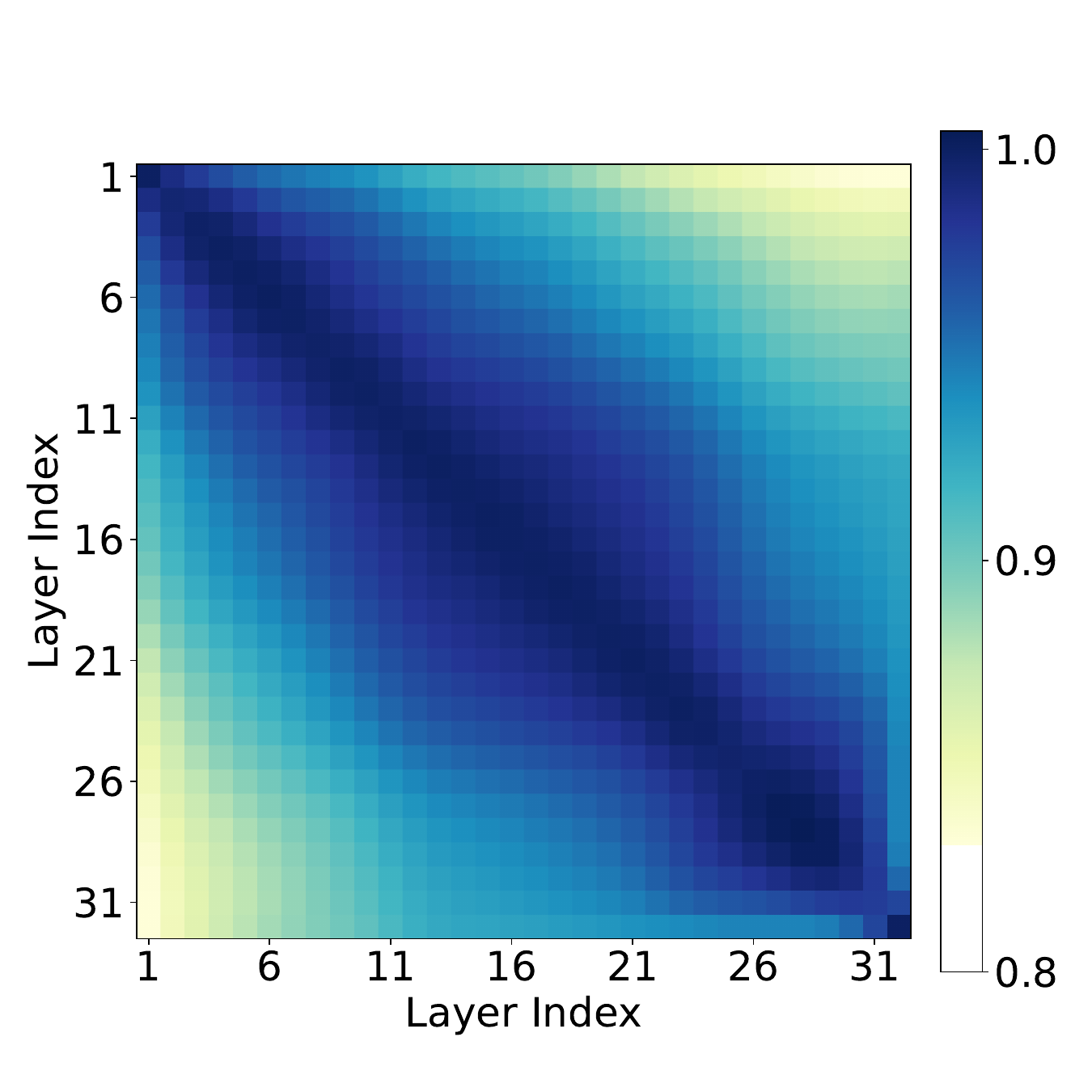}
        \caption{Layer similarity matrix at round 6.}
        \label{fig_round6}
    \end{subfigure}\hfill
    \begin{subfigure}{0.48\textwidth}
        \centering
        \includegraphics[width=\linewidth]{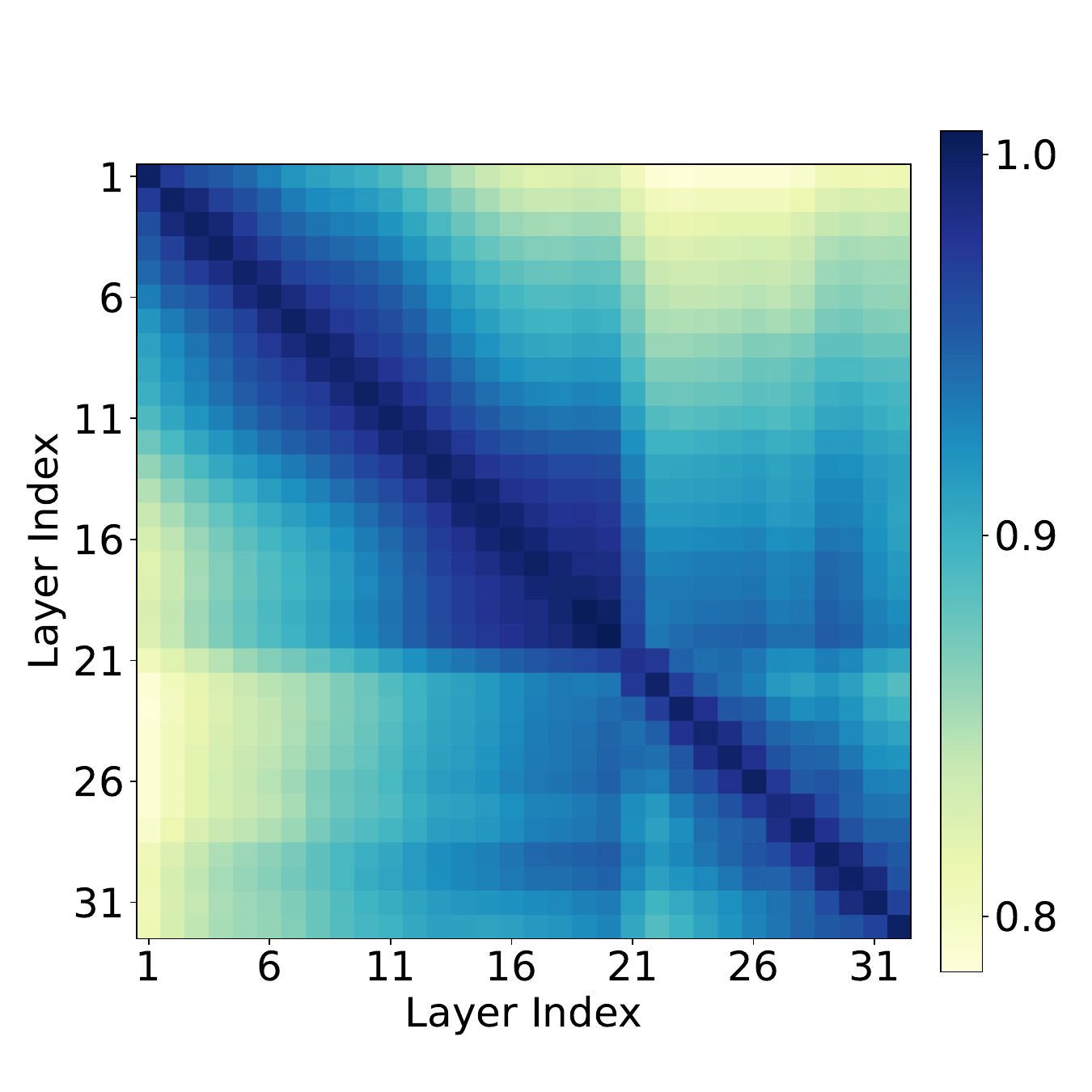}
    \caption{Layer similarity matrix at round 194.}
        \label{fig_round194}
    \end{subfigure}
    \caption{Layer similarity matrices at different rounds for the device with 3 GB available memory.}
    \label{fig_similarity}
\end{figure}

For the device with 3 GB available memory, it can accommodate the fine-tuning process of a submodel with 12 transformer layers for LLaMA2-7B (INT4). To provide deeper insights into the macro-micro synergistic pruning framework, we present the layer similarity matrices, layer grouping results, and the selection probabilities of layers within each group at two critical stages: the early phase (round 6) and the late phase (round 194) of federated fine-tuning. The grouping results and layer selection probabilities are shown in Table~\ref{table_macro_micro_round6} and Table~\ref{table_macro_micro_round194}, respectively. At round 6, we observe stronger similarities among middle layers, with $\mathcal{G}_{6}$ containing 21 layers. 
This indicates the presence of numerous functionally similar layers in the model, validating the effectiveness of our macro-level functionality-driven layer orchestration mechanism and providing opportunities for layer pruning.
As training progresses, this clustering effect gradually extends to both ends of the model, as evidenced at round 194 where $\mathcal{G}_{1}$ clusters layers 1-5, $\mathcal{G}_{6}$ clusters layers 11-19, and $\mathcal{G}_{12}$ clusters layers 27-32. This demonstrates the importance of dynamic layer grouping. Additionally, we observe significant variations in layer selection probabilities within $\mathcal{G}_{1}$ and $\mathcal{G}_{12}$, validating the effectiveness of our micro-level importance-aware layer selection strategy. Figure~\ref{fig_similarity} presents the corresponding similarity matrices, further illustrating this layer clustering phenomenon.

\section{Understanding the Fine-Grained Pruning Strategy}\label{appendix_fine_grained}

We present the layer grouping results, intra-group layer selection probabilities, and selected layers for submodel construction obtained through component-level pruning for a 3 GB device at rounds 6 and 194. 
Compared to layer-level pruning which divides the model into 12 groups, component-level pruning partitions the model into 24 groups, where one component is selected from each group for submodel construction.
The experimental results are shown in Tables~\ref{table_fine_grained_round6} and ~\ref{table_fine_grained_round194}, respectively.
Unlike layer-level pruning, we observe that component-level pruning enables more fine-grained functional grouping of the model, where MHA and FFN from the same transformer layer can be assigned to different groups, while offering greater flexibility in constructing submodels.

\input{Table/table_fine_grained_appendix1}
\input{Table/table_fine_grained_appendix2}

\section{Limitations}\label{appendix_limitation}


While \our outperforms state-of-the-art methods in terms of memory efficiency, computational requirements, communication overhead, and model performance, several limitations warrant further investigation. First, our approach only focuses on layer pruning without exploring its potential integration with complementary techniques such as knowledge distillation. Incorporating distillation mechanisms during the pruning process could potentially further enhance model performance. We leave it for future research. Second, in practical federated learning deployments, devices may dynamically join or leave the training process. Future work could explore how to more robustly coordinate the pruning process to accommodate heterogeneous and time-varying device populations while maintaining convergence guarantees and model performance.

\section{Broader Impacts}\label{appendix_broader_Impacts}

\textbf{Positive Impacts.} \our significantly reduces memory resources required for local LLM fine-tuning through intelligent layer pruning, enabling more resource-constrained devices to participate in the federated fine-tuning process, effectively utilizing their data to cultivate higher-performing models. This capability is particularly valuable in healthcare, environmental monitoring, and education sectors, especially in underserved regions where resources are scarce but the potential societal benefits of AI are substantial.

Beyond accessibility, \our promotes environmental sustainability in AI by minimizing communication overhead and computational requirements, directly reducing energy consumption and carbon emissions. This framework further strengthens privacy protection by keeping more computation on users' devices, reducing the need to centralize sensitive data and aligning with modern data protection principles.

\textbf{Negative Impacts.} As a general-purpose framework for memory-efficient federated fine-tuning, \our does not introduce negative societal impacts beyond those inherent to the underlying machine learning systems it optimizes. The technique is application-agnostic and designed to improve efficiency rather than enable new capabilities that could be misused. However, we acknowledge that any technology making AI systems more accessible and deployable could potentially amplify existing risks if applied to systems designed for harmful purposes. By lowering the resource barrier for fine-tuning language models, \our could inadvertently facilitate the deployment of AI systems without adequate safeguards in place.

We encourage practitioners to apply \our within established ethical guidelines for AI development and deployment, with appropriate consideration for fairness, accountability, and transparency. Implementers should conduct thorough impact assessments before deploying systems optimized with our framework, especially in sensitive domains or applications affecting vulnerable populations.

\end{document}

%% file: math_commands.tex
\usepackage{amsmath,amsfonts,bm}

\def\eqref#1{equation~\ref{#1}}

\def\1{\bm{1}}

\DeclareMathAlphabet{\mathsfit}{\encodingdefault}{\sfdefault}{m}{sl}
\SetMathAlphabet{\mathsfit}{bold}{\encodingdefault}{\sfdefault}{bx}{n}



%% file: Section/0_Abstract.tex
\begin{abstract}
Federated fine-tuning enables privacy-preserving Large Language Model (LLM) adaptation, but its high memory cost limits participation from resource-constrained devices. We propose \our, an innovative federated fine-tuning paradigm that tackles this via intelligent layer pruning. \our flexibly prunes the global model, creating personalized submodels based on device memory constraints. It employs a macro-micro synergistic pruning framework: a macro-level functionality-driven layer orchestration mechanism groups layers, while a micro-level importance-aware layer selection strategy prunes within groups to build device-specific submodels. We further introduce a fine-grained variant that independently prunes Multi-Head Attention and Feed-Forward Network components to precisely preserve critical architectural elements. Extensive experimental results demonstrate that \our significantly outperforms state-of-the-art approaches, achieving up to a 1.98\% improvement in average model accuracy while reducing peak memory usage by 75\%. 
\end{abstract}

%% file: Section/1_Introduction.tex
\section{Introduction}

Large Language Models (LLMs)~\citep{li2025system, huang2024swiftcoder} have demonstrated remarkable performance across a wide range of tasks. However, fine-tuning pre-trained LLMs for downstream tasks necessitates a significant amount of task-specific data, which is hard to obtain due to privacy concerns~\citep{wang2024flora, wang2025indoor}. Federated fine-tuning~\citep{wu2025survey, wu2025learning} has emerged as a promising solution, enabling the adaptation of LLMs while preserving data privacy. In response to the prohibitive nature of full parameter fine-tuning on resource-constrained devices, researchers have proposed various parameter-efficient federated fine-tuning approaches, with Low-Rank Adaptation (LoRA)~\citep{hu2021lora, tian2024hydralora} distinguishing itself through its exceptional efficiency and flexibility~\citep{guo2024selective}. Despite these advantages, federated fine-tuning with LoRA remains challenged by the high memory requirements of LLM fine-tuning~\citep{wang2024flora}, which dramatically outstrip the memory resources available in edge devices. For example, fine-tuning LLaMA2-7B~\citep{llama2} demands up to 26.9 GB of memory, whereas off-the-shelf devices typically have only 4-12 GB of memory~\citep{tian2024breaking, wu2024heterogeneity, wu2024neulite}.

To address the memory constraints, various approaches have been proposed, falling into two categories: 1) \emph{Rank Heterogeneity}, where devices are assigned different ranks based on their available resources~\citep{cho2024heterogeneous, wang2024flora}, thus reducing the trainable parameters in LoRA modules, and 2) \emph{Zeroth-Order Optimization}, which only performs forward propagation to update model parameters, removing the need to store intermediate activations~\citep{xu2023fwdllm, qin2023federated}. 
However, these methods yield limited memory savings, as both LoRA modules and activations constitute only a small fraction of the total memory footprint, with model parameters consuming the majority of the memory. 
For example, when fine-tuning LLaMA2-7B, model parameters occupy 92.8\% of memory, while LoRA modules and activations account for only 0.018\% and 7.2\%, respectively.
Therefore, compared to optimizing LoRA modules and activations, \textit{reducing model parameters is more promising to reduce memory usage.}

Building on this insight that model parameters dominate the memory footprint, we propose \our, an innovative federated fine-tuning paradigm that addresses memory constraints via intelligent layer pruning. 
Specifically, we strategically prune the layers of the global model, reducing model parameters to accommodate device memory constraints. However, a critical challenge lies in:
\begin{quote}
    \textit{How to coordinate the pruning process across devices to optimize model performance and training efficiency under resource constraints?}
\end{quote}
To address this challenge, we develop a macro-micro synergistic pruning framework that jointly considers layer functional characteristics and importance to coordinate the pruning process.
Specifically, at the macro level, we propose a functionality-driven layer orchestration mechanism that adaptively organizes layers into functional groups based on their roles in data processing. For each group, we select one layer for submodel construction while pruning the remaining layers, thereby guaranteeing that the resulting submodel maintains functional completeness and hierarchical information processing capabilities. Moreover, to select the most representative layer from each group, we propose a micro-level importance-aware layer selection strategy where the preservation probability of each layer correlates with its contribution to model performance.
This strategy effectively maintains critical computational pathways and accelerates model convergence.
Furthermore, to enable more flexible submodel construction, we propose \textsc{FedPruner}$^{+}$, a fine-grained pruning framework that performs pruning at the component level. Overall, our key contributions are summarized as follows:
\begin{itemize}
    \item We introduce \our that comprehensively explores how to intelligently perform layer pruning to address the memory constraints in federated fine-tuning. 
    \item We develop a macro-micro synergistic pruning framework to coordinate the pruning process across devices. Moreover, we propose a fine-grained variant that enables more precise pruning control at the component level.
    \item Extensive experiments on multiple benchmarks show that \our significantly outperforms existing methods while maintaining robustness under diverse memory constraints.
    
    

\end{itemize}

%% file: Section/Motivation_and_Observation.tex
\section{Preliminary and Motivation}

\subsection{Basics of LoRA }

The core idea of LoRA~\citep{hu2021lora} involves keeping the pre-trained weight matrix $\mathbf{W}_{0} \in \mathbb{R}^{U_{1}\times U_{2}}$ frozen while parameterizing its update $\mathbf{\Delta W}$ via low-rank factorization. Specifically, $\mathbf{\Delta W}$ is decomposed into the product of two trainable matrices $\mathbf{B} \in \mathbb{R}^{U_{1}\times r}$ and $\mathbf{A} \in \mathbb{R}^{r\times U_{2}}$, with $r \ll \min(U_{1}, U_{2})$.
The input is processed in parallel by $\mathbf{W}_{0}$ and $\mathbf{\Delta W}$, and their outputs are merged through element-wise addition. For a linear layer $h = \mathbf{W}_{0}x$, the modified forward propagation is formulated as: $h = \mathbf{W}_{0}x + \mathbf{B}\mathbf{A}x$.

\subsection{Memory Wall in Federated Fine-tuning}

\begin{figure*}[!t]
  \centering
  \includegraphics[width=1\linewidth]{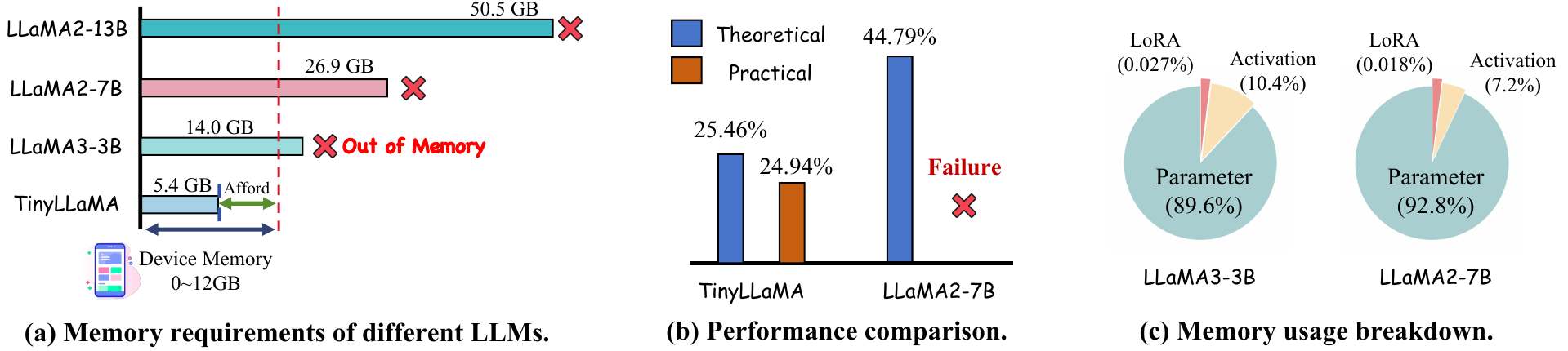}
  \caption{The memory wall in federated fine-tuning. (a) Memory requirements during fine-tuning different LLMs using LoRA. (b) Performance comparison between practical deployment and the theoretical case on the MMLU benchmark. (c) Memory usage breakdown analysis.}
  \vspace{-4mm}
  \label{fig_motivation_memorywall}
\end{figure*}

We investigate the feasibility of deploying federated fine-tuning on edge devices, focusing on memory requirements. Profiling reveals a significant ``memory wall.'' As shown in Figure~\ref{fig_motivation_memorywall}(a), fine-tuning even smaller models like TinyLLaMA with LoRA requires 5.4 GB of memory. This requirement escalates sharply for larger models such as LLaMA3-3B (14 GB) and LLaMA2-7B (26.9 GB), far exceeding the maximum available memory of 12 GB on edge devices~\citep{tam2024fedhybrid}.
Consequently, devices lack sufficient memory to perform local fine-tuning, hindering federated learning.

Moreover, this memory constraint severely impacts model performance in practical deployments. Figure~\ref{fig_motivation_memorywall}(b) demonstrates a significant performance drop compared to the theoretical scenario without memory limits. 
For TinyLLaMA~\citep{zhang2024tinyllama}, MMLU~\citep{hendrycks2020measuring} performance decreases by 0.52\%, whereas for LLaMA2-7B, the degradation is more severe as no device can support the fine-tuning process.
To pinpoint the root cause, we analyze memory allocation during LLM fine-tuning (Figure~\ref{fig_motivation_memorywall}(c)). The breakdown clearly shows that model parameters are the primary memory consumer, accounting for a dominant share (e.g., 89.6\% for LLaMA3-3B), while LoRA modules and activations consume considerably less (0.027\% and 10.4\%, respectively). This finding indicates that existing memory optimization techniques that focus on reducing LoRA parameters or activations are insufficient. Effectively breaking the memory wall requires reducing the memory consumed by the core model parameters.

\subsection{Layer Pruning to Break the Memory Wall}

\begin{wrapfigure}{r}{0.35\textwidth}
\vspace{-11pt}
\centering
\adjustbox{max width=0.35\textwidth}{\includegraphics{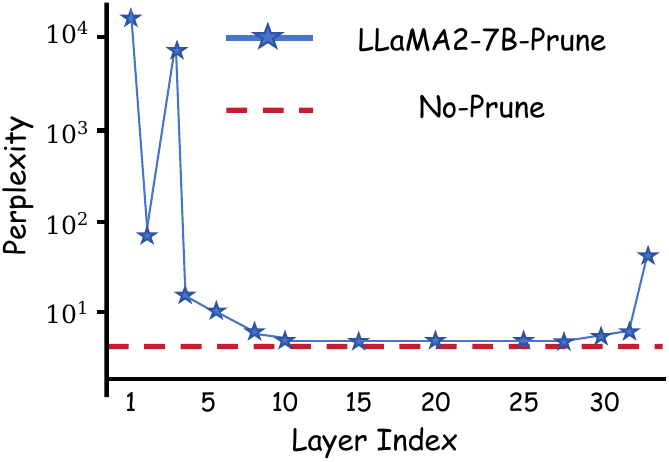}}
\vspace{-14pt}
\caption{Perplexity tested on the Alpaca-GPT4 dataset~\citep{peng2023instruction} when pruning one layer.}
\label{fig_PPL_onelayer}
\vspace{-10pt}
\end{wrapfigure}

\paragraph{Pruning Layers for Complexity Reduction.}
Motivated by the above observation, we focus on the layer pruning~\citep{men2024shortgpt} to reduce model parameters. This technique directly removes entire layers to reduce model complexity, offering inherent flexibility and hardware-agnostic advantages. 
Additionally, memory savings scale proportionally with the number of pruned layers.
We first explore whether layer pruning can decrease model complexity while maintaining model performance. To verify this hypothesis, we conduct experiments with LLaMA2-7B and evaluate the resulting perplexity (PPL) after removing individual layers. Figure~\ref{fig_PPL_onelayer} shows that significant performance degradation only occurs when removing the initial layers or the last layer, whereas pruning middle layers yields minimal performance loss.
This finding suggests that \textit{LLMs exhibit significant layer redundancy}, providing an opportunity to implement layer pruning that reduces memory consumption while maintaining model performance.


\begin{wrapfigure}{r}{0.35\textwidth}
\vspace{-11pt}
\centering
\adjustbox{max width=0.35\textwidth}{\includegraphics{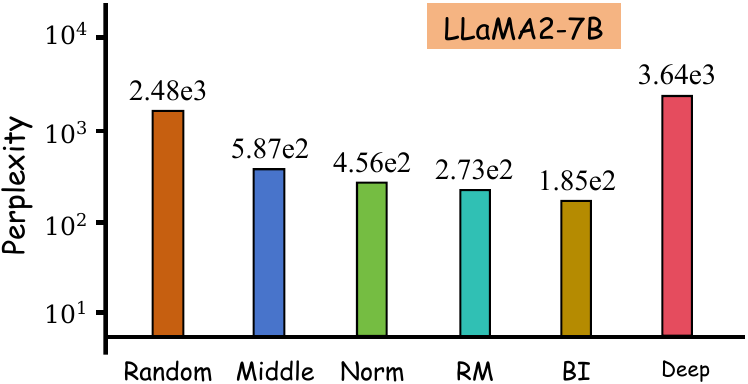}}
\vspace{-10pt}
\caption{Perplexity tested on the Alpaca-GPT4 dataset~\citep{peng2023instruction} when pruning ten layers.}
\vspace{-2mm}
\label{fig_PPL_tenlayer}
\end{wrapfigure}

\paragraph{Evaluating Multi-Layer Pruning Strategies.}
While single-layer pruning yields promising results, the stringent resource constraints of edge devices necessitate more aggressive model compression.
We therefore extend our analysis to multi-layer pruning—specifically, pruning ten layers—to further investigate the potential of layer pruning in federated fine-tuning.
We evaluate and compare six pruning strategies: 1) \textit{Random}: randomly selecting layers for removal; 
2) \textit{Middle}: pruning consecutive layers from the model's middle section; 
3) \textit{Norm}~\citep{men2024shortgpt}: pruning layers with the smallest hidden state norms;
4) \textit{Relative Magnitude (RM)}~\citep{samragh2023weight}: 
quantifying layer importance via the metric $||\frac{f(x)}{x+f(x)}||$, where $f$ represents the layer transformation function; 
5) \textit{Block Influence (BI)}~\citep{men2024shortgpt}: measuring layer significance based on the magnitude of feature modifications;
6) \textit{Deep}~\citep{gromov2024unreasonable}: prioritizing deeper layers removal due to higher knowledge retention in shallow layers. Figure~\ref{fig_PPL_tenlayer} demonstrates that when applied to multiple layer pruning scenarios, these  strategies result in substantial performance degradation. 
This deterioration highlights that fixed-rule and heuristic-based approaches become inadequate as pruning intensity increases, \textit{underscoring the need for more scientific pruning strategies to preserve model capabilities under aggressive compression.}






%% file: Section/3_FedPruner.tex
\section{Proposed Method: \our}

Drawing from these empirical findings, we introduce \our, a macro-micro synergistic pruning framework that systematically coordinates cross-device pruning by jointly considering layer functionality (Section~\ref{Macro_sec}) and layer contribution (Section~\ref{Micro_sec}). 
Furthermore, we propose a fine-grained variant (Section~\ref{sect_fine}) that independently prunes architectural components, thereby preserving critical structural elements.

\subsection{\textsc{Macro}: Functionality-Driven Layer Orchestration}\label{Macro_sec}

Naïve layer pruning induces severe functional degradation, critically undermining the submodel’s ability to capture essential features.
This stems from the hierarchical information processing of neural networks, where different layers collaboratively extract features progressing from low-level patterns to high-level semantic representations. 
To address this issue, we propose a macro-level functionality-driven layer orchestration (FDLO) mechanism. FDLO adaptively partitions layers into functional groups based on their behavioral similarity, then selects \emph{one} representative layer from each group to construct device-specific submodels. The number of groups is determined by the device’s memory budget; for instance, if a device can accommodate three layers, the entire set of layers will be divided into three functional groups.



This is based on the observation that certain layers exhibit highly similar behavioral attributes, naturally forming functional clusters. By pruning redundant layers at the group level, FDLO preserves the submodel’s hierarchical feature extraction capability while meeting device memory constraints.
Figure~\ref{fig_component1} illustrates the overview of the FDLO mechanism, which encompasses four key steps:

\begin{figure*}[!t]
  \centering
  \includegraphics[width=1\linewidth]{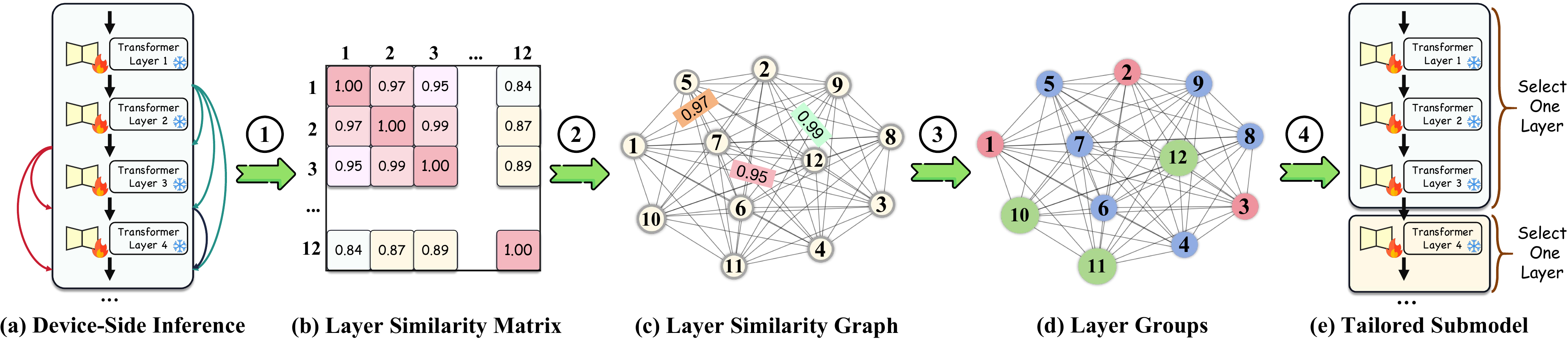}
  \caption{Overview of the functionality-driven layer orchestration mechanism, illustrated with a 12-layer Transformer model, which comprises four key steps: 1) layer similarity computation, 2) graph construction, 3) graph partitioning, and 4) submodel construction.}
  \vspace{-4mm}
  \label{fig_component1}
\end{figure*}

\textbf{\textsc{Step 1}: Layer Similarity Computation.}
Given a model with $N$ layers, each device first performs inference on its local data and computes the inter-layer output similarity using Centered Kernel Alignment (CKA)~\citep{kornblith2019similarity}, as illustrated in Figure\ref{fig_component1}(a).
Specifically, the similarity between the outputs $A_{i}$ and $A_{j}$ of layers $i$ and $j$ is computed via Equation~\ref{eq_cos}, where HSIC denotes the Hilbert–Schmidt Independence Criterion~\citep{gretton2005measuring} (see Appendix~\ref{appendix_HSIC} for its formal definition). This procedure produces a layer similarity matrix $\mathbf{W}$, exemplified in Figure~\ref{fig_component1}(b).
    \begin{equation}
    \small
    \text{CKA}(A_{i}, A_{j}) = \frac{\text{HSIC}(A_{i}, A_{j})}{\sqrt{\text{HSIC}(A_{i}, A_{i}) ~\text{HSIC}(A_{j}, A_{j})}}.
    \label{eq_cos}
    \end{equation}

\textbf{\textsc{Step 2}: Graph Construction.} 
Utilizing the computed layer similarity matrix $\mathbf{W}$, we construct a complete undirected graph $G = (\mathcal{V}, \mathcal{E})$ for each device, where vertex set $\mathcal{V}$ represents network layers and edge set $\mathcal{E}$ denotes inter-layer similarities, as shown in Figure~\ref{fig_component1}(c).

\textbf{\textsc{Step 3}: Graph Partitioning.} We then partition the constructed graph $G=(\mathcal{V}, \mathcal{E})$ into $K$ disjoint groups $\{\mathcal{G}_1,\ldots,\mathcal{G}_K\}$, where $\mathcal{G}_\phi \cap \mathcal{G}_\omega = \emptyset$ for $\phi \neq \omega$ and $\bigcup_{k=1}^K \mathcal{G}_k = \mathcal{V}$. The partitioning process consists of three key steps:
    (i) Computing the Laplacian Matrix: Given the layer similarity matrix $\mathbf{W} \in \mathbb{R}^{N \times N}$, where $w_{i,j}$ represents the pairwise similarity between layers $i$ and $j$, we construct the degree matrix $\mathbf{D} = \text{diag}(d_1,\ldots,d_N)$, where each diagonal element $d_i = \sum_{j=1}^N w_{i,j}$ represents the sum of edge weights connected to vertex $i$. The Laplacian matrix is calculated as $\mathbf{L} = \mathbf{D} - \mathbf{W}$. 

    (ii) Eigendecomposition: We perform eigenvalue decomposition on $\mathbf{L}$ and extract the $K$ smallest non-zero eigenvalues with their corresponding eigenvectors.
    These eigenvectors form the optimal low-dimensional representation of the graph, where $K$ is determined by the number of layers that the device memory can afford.
    (iii) Layer Grouping: Using the selected $K$ eigenvectors, we construct a feature matrix $\mathbf{E} \in \mathbb{R}^{N \times K}$ where each row represents a vertex as a $K$-dimensional vector. We then apply k-means to partition these vertices into $K$ groups, as shown in Figure~\ref{fig_component1}(d).
    The whole procedure can be formulated as Equation~\ref{eq_spectral_cluster}.
    \begin{equation}
    \small
    \label{eq_spectral_cluster}
    \begin{aligned}
    \{\mathcal{G}_1,...,\mathcal{G}_K\} &= \text{k-means}(\mathbf{E}\in \mathbb{R}^{N \times K}),\\
    \text{where}~\mathbf{E} &= \text{EigVectors}_{K} \Bigg(
    \text{diag}(\sum_{j=1}^N w_{i,j})_{i=1}^N-\mathbf{W}\Bigg).
    \end{aligned}
    \end{equation}
    
\textbf{\textsc{Step 4}: Submodel Construction.} Following layer grouping results $\{\mathcal{G}_1,...,\mathcal{G}_K\}$, we select \emph{one} representative layer from each group to construct a $K$-layer submodel that satisfies the device memory constraints while preserving hierarchical feature extraction capabilities, as shown in Figure~\ref{fig_component1}(e).

Additionally, to address the challenge of data heterogeneity, we group layers using locally computed similarity matrices, enabling each device to independently organize layers in a way that best matches its unique data characteristics.
Furthermore, considering that layer functionality evolves during fine-tuning, we adaptively recalculate the similarity matrices for each device to maintain synchronization between layer organizations and the model's state, thereby enhancing training stability.

\subsection{\textsc{Micro}: Importance-Aware Layer Selection}\label{Micro_sec}

While the macro-level layer orchestration mechanism effectively preserves the submodel's functional completeness, selecting the most representative layer within each group remains challenging.
This challenge stems from the fact that although layers within the same group exhibit similar functionalities, they still demonstrate notable differences in data processing. These differences become particularly pronounced when the model is divided into fewer groups, as each group inevitably encompasses a broader spectrum of layer characteristics, leading to larger intra-group variations.

To address this challenge, we propose a micro-level importance-aware layer selection (IALS) strategy that assigns retention probabilities to layers based on their contributions to model performance.
Specifically, we quantify layer contributions by analyzing input-output feature disparities, where larger disparities indicate greater feature transformations and thus higher importance. By prioritizing these high-impact layers, IALS effectively preserves critical computational paths and accelerates model convergence.
The micro-level IALS strategy involves three key steps:

\textbf{\textsc{Step 1}: Layer Importance Quantification.}  
Given the layer similarity matrix $\mathbf{W} \in \mathbb{R}^{N \times N}$, where $w_{n-1,n}$ measures the input–output feature similarity of layer $n$, we define its importance score as $\sigma_{n} = 1 - w_{n-1,n}$, capturing the magnitude of feature transformation and the layer’s contribution to the model output.

\textbf{\textsc{Step 2}: Selection Probability Generation.} 
Layer importance scores are converted to selection probabilities via a group-wise softmax.  
For layer $n$ in $\mathcal{G}_k$, the selection probability is defined as:
    \begin{equation}
    \small
        p_n = \frac{\exp(\sigma_{n})}{\sum_{m \in \mathcal{G}_k} \exp(\sigma_{m})}.
    \end{equation}

\textbf{\textsc{Step 3}: Representative Layer Selection.} For each group, \emph{one} representative layer is stochastically sampled based on selection probabilities to construct device-specific submodels, while non-selected layers are pruned. This strategy effectively exploits high-importance layers while exploring diverse layer combinations, facilitating the efficient discovery of optimal submodel architectures.

Following micro-level layer selection, each device obtains its customized submodel and conducts local fine-tuning. 
Only the updated LoRA parameters are then transmitted to the central server for aggregation, as detailed in Appendix~\ref{appendix_aggregation}.
The complete workflow of \our, encompassing both macro-level orchestration and micro-level selection, is described in Appendix~\ref{appendix_workflow}. Furthermore, we provide the convergence analysis of \our in Appendix~\ref{sec:convergence_proof}.

\begin{wrapfigure}{r}{0.45\textwidth}
\vspace{-15pt}
\centering
\adjustbox{max width=0.45\textwidth}{\includegraphics{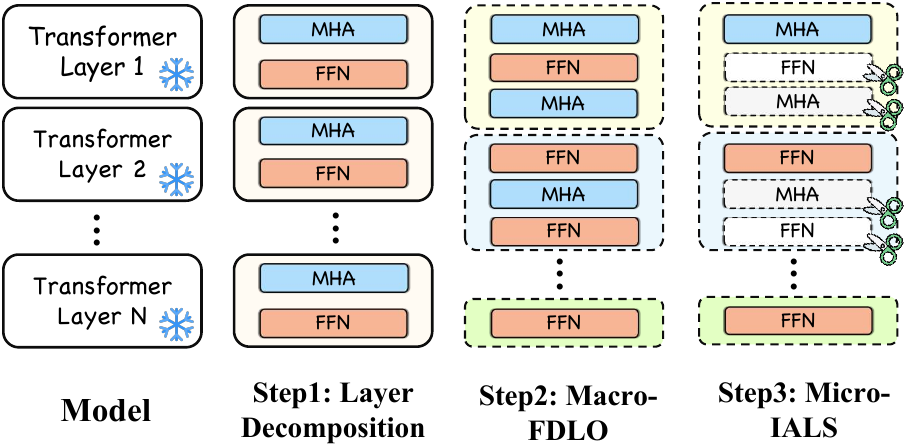}}
\vspace{-10pt}
\caption{Workflow of the \textsc{FedPruner}$^{+}$.}
\label{FedPruner_component3}
\vspace{-4mm}
\end{wrapfigure}

\subsection{\textsc{Fine-Grained}: Component-Level Pruning}\label{sect_fine}

In this section, we present \textsc{FedPruner}$^{+}$, a fine-grained variant of \our that extends the macro–micro synergistic pruning framework from the layer level to the component level.
This design builds on two key observations: 
1) \textbf{Modular Independence}: Each transformer layer consists of two primary components, Multi-Head Attention (MHA) and Feed-Forward Network (FFN), which operate independently while maintaining dimensional consistency. This enables selective component pruning without introducing dimensional mismatches.
2) \textbf{Heterogeneous Functionality}: MHA and FFN serve distinct roles, with MHA capturing contextual relationships and FFN performing non-linear  transformations. These observations motivate us to treat them as independent prunable units.

The workflow of \textsc{FedPruner}$^{+}$, illustrated in Figure~\ref{FedPruner_component3}, differs from \our in requiring layer decomposition before executing macro-micro synergistic pruning, and consists of three steps:
1) \textbf{Layer Decomposition}: Each layer is decomposed into its constituent MHA and FFN components, which are treated as independent prunable units. 
2) \textbf{\textsc{Macro}-FDLO}: Macro-level layer orchestration is performed on the decomposed architecture, producing 2$K$ groups.
3) \textbf{\textsc{Micro}-IALS}: Within each group, micro-level selection identifies \emph{one} representative component for submodel construction. 
In this way, we can achieve more precise retention of critical architectural elements.

%% file: Section/4_Experiments.tex
\section{Experiments}

\subsection{Experimental Setup}\label{sec_setup}


Consistent with OpenFedLLM~\citep{ye2024openfedllm}, we evaluate \our and its fine-grained variant on instruction tuning tasks across three LLaMA-based LLMs with varying parameter scales: TinyLLaMA (1.1B)~\citep{zhang2024tinyllama}, LLaMA2-7B, and LLaMA2-13B~\citep{llama2}. All models are fine-tuned on the Alpaca-GPT4 dataset~\citep{peng2023instruction}.  
Evaluation is conducted on both close-ended and open-ended benchmarks. The close-ended evaluation suite includes TruthfulQA~\citep{lin2022truthfulqa} (truthfulness), MMLU~\citep{hendrycks2020measuring} (knowledge), IFEval~\citep{zhou2023instructionfollowing} (instruction following), and BBH~\citep{bbh} (reasoning). The open-ended evaluation adopts Vicuna-Bench~\citep{chiang2023vicuna} and MT-Bench~\citep{zheng2024judging} to assess multi-turn conversational ability. Additional experimental details are provided in Appendix~\ref{appendix_setup}.

\subsection{Baselines}


We compare \our with a theoretical baseline (LoRA), which assumes all devices have sufficient memory to support local fine-tuning, and report Zero-Shot performance as a lower bound. We further benchmark against two categories of methods: 1) \textbf{Memory-Unaware Methods}: FedIT~\citep{zhang2024towards}, DoFIT~\citep{xu2024dofit}, and FedSA-LoRA~\citep{guo2024selective}; 2) \textbf{Memory-Aware Methods}: FLoRA~\citep{wang2024flora}, FwdLLM~\citep{xu2023fwdllm}, and FedRA~\citep{su2024fedra}. Detailed baseline descriptions are provided in Appendix~\ref{appendix_baseline}.

\subsection{Overall Evaluation}

\input{Table/Table_Main}

Table~\ref{tab_overall_performance} presents the experimental results, which demonstrate that both \our and its fine-grained variant consistently outperform baseline methods across all evaluation benchmarks. 


1) \textbf{Comparison with Memory-Unaware Methods.} As model size grows, device participation rates for these methods drop sharply—from 60\% on TinyLLaMA to 15\% on LLaMA2-7B, and to 0\% on LLaMA2-13B—causing substantial performance loss. For instance, on close-ended benchmarks, FedIT exhibits a 2.07\% average performance degradation on TinyLLaMA, which widens to 5.42\% on LLaMA2-7B and ultimately reaches 11.11\% on LLaMA2-13B compared to \our. DoFIT and FedSA-LoRA display similar patterns of performance deterioration, further underscoring the scalability limitations of memory-unaware approaches.


2) \textbf{Comparison with Memory-Aware Methods.} 
FLoRA optimizes only LoRA modules, yielding limited memory savings and thus showing performance degradation similar to memory-unaware methods.
FwdLLM achieves moderate memory optimization by eliminating the 
need to store intermediate activations, increasing device participation from 60\% to 75\% (TinyLLaMA) and 15\% to 45\% (LLaMA2-7B), but still fails on  LLaMA2-13B. 
FedRA enables local fine-tuning under memory constraints but lacks a systematic layer allocation strategy, leading to a 4.3\% (close-ended) and 0.46 (open-ended) average performance drop on LLaMA2-13B compared to \our.

3) \textbf{Comparison with Theoretical Baseline.} \our consistently outperforms the theoretical baseline (LoRA). For instance, on LLaMA2-13B, it delivers average performance gains of 3.5\% on close-ended benchmarks and 0.7 on open-ended benchmarks, while reducing memory consumption by 58.5\%. These gains arise from \our's intelligent layer pruning strategy, which selectively fine-tunes the most critical layers with minimal disruption to the model's pre-trained knowledge.



4) \textbf{Benefits of Fine-Grained Pruning.} \textsc{FedPruner}$^{+}$ outperforms \our across all evaluation settings, with gains amplifying as model complexity increases. For example, on close-ended benchmarks, the average performance improvement rises from 0.26\% for TinyLLaMA to 0.47\% for LLaMA2-7B, and 0.88\% for LLaMA2-13B. These gains stem from its fine-grained pruning strategy, which more effectively preserves essential architectural components.

\subsection{Understanding the Macro-Micro Synergistic Pruning Framework}

\begin{wrapfigure}{r}{0.5\textwidth}
\vspace{-15pt}
\centering
\adjustbox{max width=0.5\textwidth}{\input{Graph/figure_macro_micro_impact}}
\vspace{-6mm}
\caption{Impact of the macro-micro synergistic pruning framework: (a) MMLU accuracy and (b) layer selection frequency of constructed submodels for a device with 3 GB memory.}
\vspace{-3mm}
\label{fig:macro_micro}
\end{wrapfigure}

To better understand the macro-micro synergistic pruning framework, we present a detailed analysis focusing on three critical dimensions: model performance, convergence efficiency, and layer selection patterns. Specifically, we experiment with LLaMA2-7B and compare with an ablated variant (denoted as w/o MM) that omits the synergistic pruning framework. 
As shown in Figure~\ref{fig:macro_micro}(a), \our improves MMLU accuracy by 1.23\% and converges 1.33$\times$ faster compared to w/o MM. 
To elucidate the fundamental reasons driving this performance enhancement, Figure~\ref{fig:macro_micro}(b) illustrates the layer selection patterns during submodel construction for a 3 GB device throughout the federated fine-tuning process. We observe that \our strategically prioritizes head and tail layers in the constructed submodels, diverging from uniform sampling strategy. This scientific submodel construction method effectively improves model performance and training efficiency. Moreover, this observed layer selection pattern aligns with our previous analysis, suggesting a higher degree of redundancy in intermediate layers. More analyses are presented in Appendix~\ref{appendix_Macro_Micro}.

\subsection{Understanding the Fine-Grained Pruning Strategy}

\vspace{-2mm}
\begin{figure*}[!h]
  \centering
  \includegraphics[width=0.85\linewidth]{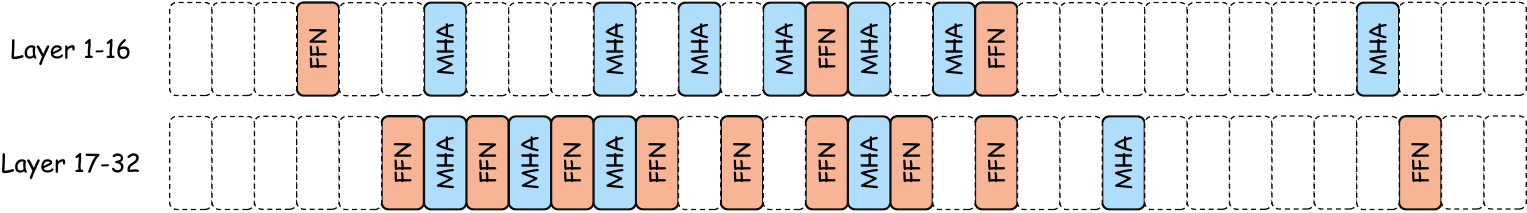}
    \caption{Architecture of the submodel constructed through component-level pruning.}
  \vspace{-2mm}
  \label{fig_fine_grained_model}
\end{figure*}

In this section, we investigate the key factors contributing to the success of our fine-grained pruning strategy by analyzing the composition of constructed submodels. Figure~\ref{fig_fine_grained_model} illustrates the constructed submodel for a device with 3 GB memory at round 194, with LLaMA2-7B serving as the global model. We observe that component-level pruning 
enables more flexible architectural configurations. 
For example, the FFN from layer 2 connects to the MHA from layer 4, indicating that pairing MHA and FFN within the same layer is not always optimal. Notably, the MHA components of layers 4, 6, 7, and 8 are directly connected, suggesting that the conventional alternating MHA–FFN pattern may be suboptimal. \textsc{FedPruner}$^{+}$ leverages its flexible pruning strategy to enable optimized architectural arrangements, thereby achieving superior performance. Additional experimental results and analysis are presented in Appendix~\ref{appendix_fine_grained}.

\subsection{Performance Under Different Memory Constraints}

\begin{wraptable}{R}{0.55\textwidth}
\vspace{-12pt}
\centering
\caption{Performance under different memory constraints. The percentage indicates the proportion of global model size each device can handle.}
\label{table_differnet_memory_constraints}
\input{Table/table_different_memory_constraints}
\vspace{-4mm}
\end{wraptable}
In this section, we evaluate \our's effectiveness under different memory constraints. Specifically, we configure uniform memory constraints across all devices, varying the memory capacity from 25\% to 75\% of the global model size. For instance, 25\% indicates that devices can only accommodate one-fourth of the global model size during local fine-tuning. We benchmark against FedRA as it represents the most closely related work to our approach.
Table~\ref{table_differnet_memory_constraints} presents the experimental results, demonstrating that \our consistently outperforms FedRA across all settings. For LLaMA2-7B, \our delivers an average performance improvement of 3.10\% at 25\% memory budget, 2.54\% at 50\%, and 2.17\% at 75\% compared to FedRA. This performance enhancement becomes even more pronounced on LLaMA2-13B. Specifically, under a 25\% memory constraint on LLaMA2-13B, \our achieves a remarkable 5.57\% average accuracy improvement compared to FedRA and even surpasses LoRA by 1.98\% while reducing peak memory usage by 75\%.
These compelling results substantiate the robustness of \our under diverse memory constraints and the effectiveness of its pruning strategy in parameter optimization.

\subsection{Ablation Study}

We further conduct extensive ablation studies to evaluate the contributions of each technique proposed in \our, i.e., functionality-driven layer orchestration (FDLO) and importance-aware layer selection (IALS). 
The experimental results presented in Table~\ref{ablation_study} demonstrate that both techniques significantly enhance the model performance, with their benefits becoming more pronounced as model complexity increases. For LLaMA2-7B, removing FDLO and IALS leads to average performance degradation of 1.22\% and 0.57\% respectively, while removing both results in a 1.71\% decline. The impact is more substantial on LLaMA2-13B, where the performance drops reach 2.85\% and 1.29\% for individual removals, and 4.3\% when both techniques are disabled. These results validate the effectiveness of each technique and the complementary benefits of our proposed macro-micro synergistic pruning framework. Moreover, this empirical evidence further explains that performance degradation in layer importance-based pruning methods primarily stems from their inability to preserve the hierarchical information processing capabilities of the resulting submodels.


\begin{figure}[!t]
    \centering
    \vspace{-4mm}
    \begin{minipage}[c]{0.55\textwidth}
        \centering
        \begin{table}[H]
        \caption{Ablation study of \our.}
        \label{ablation_study}
        \resizebox{\linewidth}{!}{
        \begin{tabular}{lccccc}
        \toprule[1pt]
        \multirow{2}{*}{\textbf{Method}}  & \multicolumn{4}{c}{\textbf{Close-Ended Benchmark}} &\multirow{2}{*}{\textbf{Average}} \\  
        
        \cmidrule{2-5}
        
        & \textbf{TruthfulQA} & \textbf{MMLU} & \textbf{IFEval} & \textbf{BBH}   \\ 
        \midrule[1pt]
        &\multicolumn{4}{c}{\textbf{LLaMA2-7B (INT4)}} &\\ 
        \midrule[1pt]
        \rowcolor{my_c1!70}
        \our & \textbf{49.95} & \textbf{43.64} & \textbf{33.72} & \textbf{40.65} & \textbf{41.99} \\
        w/o FDLO                    & 48.14 & 42.93 & 32.44 & 39.58 & 40.77 (-1.22\%) \\ 
        w/o IALS                    & 49.38 & 43.02 & 32.87 & 40.42 & 41.42 (-0.57\%) \\
        w/o FDLO \& IALS & 47.66 & 42.41 & 31.67 & 39.36 & 40.28 (-1.71\%) \\
        
        \midrule[1pt]
        &\multicolumn{4}{c}{\textbf{LLaMA2-13B (INT4)}} &\\ 
        \midrule[1pt]
        \rowcolor{my_c1!70}
        \our & \textbf{56.13} & \textbf{58.09} & \textbf{45.62} & \textbf{48.46} & \textbf{52.08} \\
        w/o FDLO                    & 53.24 & 55.76 & 41.47 & 46.45 & 49.23 (-2.85\%) \\ 
        w/o IALS                    & 54.82 & 56.94 & 43.64 & 47.74 & 50.79 (-1.29\%) \\
        w/o FDLO \& IALS & 51.69 & 54.93 & 39.11 & 45.39 & 47.78 (-4.30\%) \\
        \bottomrule[1pt]
        \end{tabular}
        }
        \end{table}
    \end{minipage}
\hfill 
\begin{minipage}[c]{0.4\textwidth}
    \centering
    \begin{table}[H]
    \caption{Overhead analysis of various methods on LLaMA2-7B.}
    \label{table_overhead_analysis}
    \renewcommand{\arraystretch}{0.9} 

    \resizebox{\linewidth}{!}{
    \begin{tabular}{lccccc}
    \toprule[1pt]
    \multirow{2}{*}{\textbf{Method}} & \multicolumn{2}{c}{\textbf{Resource Consumption ($\downarrow$)}}\\  
    
    \cmidrule{2-3}
    
    & \textbf{Time (h)} & \textbf{Communication (GB)} \\ 
    \midrule[1pt]
    FedIT & 3.22 & 6.05 \\
    DoFIT & 4.48 & 7.62\\
    FedSA-LoRA & 2.93 & 5.46\\
    FLoRA & 3.61 & 7.46\\
    FwdLLM & 2.98 & 5.58\\
    FedRA & 2.10 & 3.65 \\
    LoRA & 2.56 & 5.03\\
    \midrule[0.5pt]
    \rowcolor{my_c1!50}
    \our & \textbf{1.92} & \textbf{3.47} \\
    \bottomrule[1pt]
    \end{tabular}
    }
    \end{table}
\end{minipage}
\vspace{-5mm}
\end{figure}

\subsection{Overhead Analysis}


Finally, we evaluate \our's efficiency in terms of convergence time and communication cost using LLaMA2-7B as the global model on NVIDIA H800 GPUs. 
As shown in Table~\ref{table_overhead_analysis}, \our surpasses all baselines, achieving up to $2.33\times$ faster convergence and $2.20\times$ lower communication cost. 
These substantial gains stem from its macro–micro synergistic pruning framework, which systematically coordinates cross-device pruning to accelerate model convergence. 
Overall, \our is a resource-efficient approach, making it particularly well-suited for federated fine-tuning of LLMs in resource-constrained environments.

%% file: Table/Table_Main.tex
\definecolor{steelbluev2}{HTML}{DAE8FC}
\definecolor{steelblue}{HTML}{82B0D2}
\definecolor{my_c1}{HTML}{aacfd0}
\definecolor{my_c2}{HTML}{e4eddb}

\begin{table*}[t]


\caption{Performance evaluation on instruction tuning tasks. \textbf{Bold} and \underline{underlined} values denote the best and second-best performance, respectively. The symbol \colorbox{green!30}{``$-$''} indicates that the method is not applicable due to the lack of devices with sufficient memory resources to support local fine-tuning, thereby degenerating to Zero-Shot. The \colorbox{orange!30}{$PR$} represents the device participation rate.}

\small 
\renewcommand{\arraystretch}{0.95} 
  \centering
  \resizebox{1\textwidth}{!}{ 
  
    \begin{tabular}{llcccccccccc}
    \toprule[1pt]
    &\multirow{2}{*}{\textbf{Method}}&\multicolumn{4}{c}{\textbf{Close-Ended Benchmark}} &\multirow{2}{*}{\textbf{Average}}&
    \multicolumn{3}{c}{\textbf{Open-Ended Benchmark}}&  \multirow{2}{*}{\textbf{Average}} &  \multirow{2}{*}{\colorbox{orange!30}{\textbf{PR}}} \\
    \cmidrule{3-6}
    \cmidrule{8-10}
    && TruthfulQA & MMLU & IFEval & BBH & & Vicuna & MT-1 & MT-2 \\

    \midrule[1pt]
    &&\multicolumn{10}{c}{\textbf{TinyLLaMA}~\citep{zhang2024tinyllama}} \\ 
    \midrule[1pt]
    \rowcolor{gray!20}
    & Zero-Shot & 37.58 & 24.72 & 16.05 & 25.91 & 26.07 & 5.52 & 1.61 & 1.18 & 2.77 & $/$ \\
    \midrule[0.5pt]

    \multirow{3}{*}{\textbf{\shortstack{Memory\\Unaware}}}& FedIT      & 37.87 & 24.94 & 16.19 & 26.03 & 26.26 & 5.68 & 1.78 & 1.29 & 2.92 & 60\% \\
    &DoFIT           & 38.81 & 25.64 & 18.72 & 27.42 & 27.65 & 6.13 & 2.27 & 1.44 & 3.28 & 60\% \\
    &FedSA-LoRA & 38.63 & 25.56 & 19.01 & 27.51 & 27.68 & 6.25 & 2.35 & 1.51 & 3.37 & 60\% \\
    \midrule[0.5pt]
    \multirow{5}{*}{\textbf{\shortstack{Memory\\Aware}}}&FLoRA        & 38.47 & 25.35 & 17.58 & 27.24 & 27.16 & 5.84 & 1.96 & 1.33 & 3.04 & 60\% \\
    &FwdLLM        & 38.55 & 25.49 & 18.64 & 27.35 & 27.51 & 5.98 & 2.09 & 1.36 & 3.14 & 75\% \\ 
    &FedRA          & 38.74 & 25.52 & 19.23 & 27.57 & 27.77 & 6.26 & 2.40 & 1.57 & 3.41 & 100\% \\
    &\cellcolor{my_c1!50}\our               & \cellcolor{my_c1!50}\underline{39.16} & \cellcolor{my_c1!50}\underline{26.18} & \cellcolor{my_c1!50}\underline{19.87} & \cellcolor{my_c1!50}\underline{28.09} & \cellcolor{my_c1!50}\underline{28.33} & \cellcolor{my_c1!50}\underline{6.87} & \cellcolor{my_c1!50}\underline{2.81} & \cellcolor{my_c1!50}\underline{1.92} & \cellcolor{my_c1!50}\underline{3.87} & \cellcolor{my_c1!50}100\% \\  
    &\cellcolor{my_c1!50}\textsc{FedPruner}$^{+}$     & \cellcolor{my_c1!50}\textbf{39.32} & \cellcolor{my_c1!50}\textbf{26.42} & \cellcolor{my_c1!50}\textbf{20.26} & \cellcolor{my_c1!50}\textbf{28.35} & \cellcolor{my_c1!50}\textbf{28.59} & \cellcolor{my_c1!50}\textbf{7.00} & \cellcolor{my_c1!50}\textbf{2.94} & \cellcolor{my_c1!50}\textbf{2.01} & \cellcolor{my_c1!50}\textbf{3.98} & \cellcolor{my_c1!50}100\% \\ 
    \midrule
    \rowcolor{my_c2!50}
    \textbf{Theoretical}&LoRA             & 38.58 & 25.46 & 18.67 & 27.33 & 27.51 & 6.02 & 2.13 & 1.35 & 3.17 & 100\% \\

    \midrule[1pt]
    &&\multicolumn{10}{c}{\textbf{LLaMA2-7B (INT4)}~\citep{llama2}} \\ 
    \midrule[1pt]
    \rowcolor{gray!20}

    & Zero-Shot & 40.98 & 40.20 & 25.46 & 36.95 & 35.90 & 7.00 & 3.04 & 1.13 & 3.72 & $/$ \\
    \midrule[0.5pt]
    \multirow{3}{*}{\textbf{\shortstack{Memory\\Unaware}}}&FedIT      & 41.96 & 40.82 & 26.38 & 37.13 & 36.57 & 7.20 & 3.37 & 1.35 & 3.97 & 15\% \\
    &DoFIT           & 43.74 & 41.53 & 28.71 & 38.45 & 38.11 & 7.79 & 4.06 & 1.59 & 4.48 & 15\% \\
    &FedSA-LoRA & 45.27 & 41.77 & 29.60 & 38.62 & 38.82 & 8.03 & 4.28 & 1.82 & 4.71 & 15\% \\
    \midrule[0.5pt]
    \multirow{5}{*}{\textbf{\shortstack{Memory\\Aware}}}&FLoRA         & 43.58 & 41.35 & 28.06 & 37.94 & 37.73 & 7.57 & 3.85 & 1.47 & 4.30 & 15\% \\
    &FwdLLM         & 45.83 & 42.20 & 30.57 & 38.97 & 39.39 & 8.16 & 4.71 & 2.05 & 4.97 & 45\% \\ 
    &FedRA           & 47.66 & 42.41 & 31.67 & 39.36 & 40.28 & 8.21 & 5.12 & 2.27 & 5.20 & 100\% \\
    &\cellcolor{my_c1!50}\our              & \cellcolor{my_c1!50}\underline{49.95} & \cellcolor{my_c1!50}\underline{43.64} & \cellcolor{my_c1!50}\underline{33.72} & \cellcolor{my_c1!50}\underline{40.65} & \cellcolor{my_c1!50}\underline{41.99} & \cellcolor{my_c1!50}\underline{8.32} & \cellcolor{my_c1!50}\underline{5.64} & \cellcolor{my_c1!50}\underline{2.81} & \cellcolor{my_c1!50}\underline{5.59} & \cellcolor{my_c1!50}100\% \\
    &\cellcolor{my_c1!50}\textsc{FedPruner}$^{+}$     & \cellcolor{my_c1!50}\textbf{50.40} & \cellcolor{my_c1!50}\textbf{44.26} & \cellcolor{my_c1!50}\textbf{34.17} & \cellcolor{my_c1!50}\textbf{40.99} & \cellcolor{my_c1!50}\textbf{42.46} &\cellcolor{my_c1!50}\textbf{8.44} & \cellcolor{my_c1!50}\textbf{5.83} & \cellcolor{my_c1!50}\textbf{2.96} & \cellcolor{my_c1!50}\textbf{5.74} & \cellcolor{my_c1!50}100\% \\
    \midrule
    \rowcolor{my_c2!50}
    \textbf{Theoretical}&LoRA             & 47.57 & 42.45 & 31.76 & 39.28 & 40.27 & 8.18 & 4.77 & 1.98 & 4.98 & 100\% \\

    \midrule[1pt]
    &&\multicolumn{10}{c}{\textbf{LLaMA2-13B (INT4)}~\citep{llama2}} \\ 
    \midrule[1pt]
    \rowcolor{gray!20}
    & Zero-Shot & 42.83 & 49.65 & 30.35 & 41.03 & 40.97 & 7.27 & 3.65 & 2.07 & 4.33 & $/$  \\
    \midrule[0.5pt]
    
    \multirow{3}{*}{\textbf{\shortstack{Memory\\Unaware}}}&FedIT      & --- & --- & --- & --- & --- & --- & --- & --- & --- & 0\% \\
    &DoFIT           & --- & --- & --- & --- & --- & --- & --- & --- & --- & 0\% \\
    &FedSA-LoRA & --- & --- & --- & --- & --- & --- & --- & --- & --- & 0\% \\
    \midrule[0.5pt]
    \multirow{5}{*}{\textbf{\shortstack{Memory\\Aware}}}&FLoRA         & --- & --- & --- & --- & --- & --- & --- & --- & --- & 0\% \\
    &FwdLLM         & --- & --- & --- & --- & --- & --- & --- & --- & --- & 0\% \\ 
    &FedRA           & 51.69 & 54.93 & 39.11  & 45.39 & 47.78 & 8.46 & 5.54 & 3.28 & 5.76 & 100\% \\
    &\cellcolor{my_c1!50}\our               & \cellcolor{my_c1!50}\underline{56.13} & \cellcolor{my_c1!50}\underline{58.09} & \cellcolor{my_c1!50}\underline{45.62} & \cellcolor{my_c1!50}\underline{48.46} & \cellcolor{my_c1!50}\underline{52.08} & \cellcolor{my_c1!50}\underline{8.59} & \cellcolor{my_c1!50}\underline{5.92} & \cellcolor{my_c1!50}\underline{4.16} & \cellcolor{my_c1!50}\underline{6.22} & \cellcolor{my_c1!50}100\% \\
    &\cellcolor{my_c1!50}\textsc{FedPruner}$^{+}$      & \cellcolor{my_c1!50}\textbf{57.36} & \cellcolor{my_c1!50}\textbf{58.89} & \cellcolor{my_c1!50}\textbf{46.65} & \cellcolor{my_c1!50}\textbf{48.93} & \cellcolor{my_c1!50}\textbf{52.96} & \cellcolor{my_c1!50}\textbf{8.75} & \cellcolor{my_c1!50}\textbf{6.16} & \cellcolor{my_c1!50}\textbf{4.45} & \cellcolor{my_c1!50}\textbf{6.45} & \cellcolor{my_c1!50}100\% \\
    \midrule
    \rowcolor{my_c2!50}
    \textbf{Theoretical}&LoRA             & 52.40 & 55.45 & 40.33 & 46.14 & 48.58 & 8.37 & 5.17 & 3.01 & 5.52 & 100\% \\
    \bottomrule[1pt]
    \end{tabular}%
    }
  
\label{tab_overall_performance}
\vspace{-5mm}
\end{table*}

%% file: Graph/figure_macro_micro_impact.tex
\definecolor{Maroon}{HTML}{7CCD7C}
\definecolor{BLUE}{HTML}{8EE5EE}
\definecolor{my-green}{HTML}{8ECFC9}
\definecolor{my-yellow}{HTML}{FFBE7A}
\definecolor{my-blue}{HTML}{82B0D2}

\definecolor{my_c1}{HTML}{f19584}
\definecolor{my_c2}{HTML}{307672}

\pgfplotsset{width=0.7\linewidth,height=0.55\linewidth,compat=1.15}
\footnotesize
\begin{tikzpicture}
\scriptsize{
\begin{axis}[
	at={(0em,0em)},
    xlabel={Round},
    ylabel={Accuracy (\%)},
    xmin=0.05, xmax=0.55,
    ymin=39, ymax=45,
    xtick={0.1, 0.2, 0.3, 0.4, 0.5},
    ytick={40, 42, 44, 46},
    ymajorgrids=true,
    xmajorgrids=true,
    grid style=dashed,
    xticklabels={0, 50, 100, 150, 200},
    x label style={at={(axis description cs:0.5,-0.15)},anchor=north, font=\small},
    y label style={at={(axis description cs:-0.1,0.5)},anchor=south, font=\small},
    legend style={
    	at={(1.1,1.06)},
    	anchor=south,
    	legend columns=-1,
    	nodes={scale=1, transform shape}}
]
\addplot[
    color=my_c1,
    mark=diamond*,
    mark size=2.pt,thick,line width=2.2pt,
    mark options={fill=my_c1,draw=my_c1,line width=2.2pt}
    ]
    coordinates {
   (0.1, 40.12)
    (0.2, 41.24)
    (0.3, 41.83)
    (0.4, 42.11)
    (0.5, 42.30)
    };
    \addlegendentry{w/o MM}
\addplot[
    color=my_c2,
    mark=pentagon*,
    mark size=2.pt,thick,line width=2.2pt,
    mark options={fill=my_c2,draw=my_c2,line width=2.2pt}
    ]
    coordinates {
    (0.1, 40.12)
    (0.2, 42.10)
    (0.3, 43.15)
    (0.4, 43.60)
    (0.5, 43.64)
    };
    \addlegendentry{\our}
\end{axis}

\begin{axis}[
    at={(18em,0em)},
    xlabel={Layer Index},
    ylabel={Frequency},
    xmin=0.00, xmax=3.3,
    ymin=0.2, ymax=20,
    xtick={0.1, 0.2, 0.3, 0.4, 0.5,0.6,0.7,0.8,0.9,1,1.1,1.2,1.3,1.4,1.5,1.6,1.7,1.8,1.9,2.0,2.1,2.2,2.3,2.4,2.5,2.6,2.7,2.8,2.9,3.0,3.1,3.2},
    ytick={5,10,15,20},
    ymajorgrids=true,
    xmajorgrids=true,
    grid style=dashed,
    xticklabels={,,,4,,,,8,,,,12,,,,16,,,,20,,,,24,,,,28,,,,32},
    x label style={at={(axis description cs:0.5,-0.15)},anchor=north, font=\small},
    y label style={at={(axis description cs:-0.1,0.5)},anchor=south, font=\small}
]
\addplot[
    color=my_c1,
    mark=diamond*,
    mark size=1.5pt,thick,line width=1.2pt,
    mark options={fill=my_c1,draw=my_c1,line width=1.2pt}
    ]
    coordinates {
    (0.1, 4)
    (0.2, 4)
    (0.4, 5)
    (0.6, 3)
    (0.8, 4)
    (1.0, 5)
    (1.2, 4)
    (1.4, 3)
    (1.6, 4)
    (1.8, 4)
    (2.0, 5)
    (2.2, 4)
    (2.4, 4)
    (2.6, 3)
    (2.8, 4)
    (3.1, 4)
    (3.2, 5)

    };
\addplot[
    color=my_c2,
    mark=pentagon*,
    mark size=1.5pt,thick,line width=1.2pt,
    mark options={fill=my_c2,draw=my_c2,line width=1.2pt}
    ]
    coordinates {
    (0.1, 19)
    (0.2, 12)
    (0.3, 7)
    (0.5, 2)
    (0.7, 2)
    (0.9, 2)
    (1.1, 4)
    (1.3, 1)
    (1.5, 2)
    (1.7, 2)
    (1.9, 4)
    (2.1, 1)
    (2.3, 3)
    (2.5, 1)
    (2.7, 4)
    (3.0, 5)
    (3.1, 11)
    (3.2, 17)
    };

\end{axis}
}
\end{tikzpicture}

%% file: Table/table_different_memory_constraints.tex
\definecolor{my_c1}{HTML}{aacfd0}
\definecolor{my_c2}{HTML}{e4eddb}

\centering
\resizebox{\linewidth}{!}{
\begin{tabular}{ccccccc}
\toprule[1pt]
{\textbf{\shortstack{Memory}}}&\multirow{2}{*}{\textbf{Method}}  & \multicolumn{4}{c}{\textbf{Close-Ended Benchmark}} &\multirow{2}{*}{\textbf{Average}} \\  

\cmidrule{3-6}

\textbf{Constraints}&& \textbf{TruthfulQA} & \textbf{MMLU} & \textbf{IFEval} & \textbf{BBH}   \\ 

\midrule[1pt]
&&\multicolumn{4}{c}{\textbf{LLaMA2-7B (INT4)}} &\\ 
\midrule[1pt]
\multirow{2}{*}{\textbf{25\%}} &FedRA                        & 42.54 & 41.03 & 27.72 & 37.45 & 37.19 \\ 
                               & \our & \textbf{46.85} & \textbf{42.67} & \textbf{31.96} & \textbf{39.66} & \colorbox{my_c1!50}{\textbf{40.29} (+3.10\%)} \\

\midrule[0.5pt]

\multirow{2}{*}{\textbf{50\%}} &FedRA                        & 44.93 & 41.62 & 29.57 & 38.34 & 38.62 \\ 
                               & \our & \textbf{48.67} & \textbf{43.16} & \textbf{32.74} & \textbf{40.05} & \colorbox{my_c1!50}{\textbf{41.16} (+2.54\%)} \\

\midrule[0.5pt]

\multirow{2}{*}{\textbf{75\%}} &FedRA                        & 46.58 & 42.14 & 30.86 & 39.02 & 39.65 \\ 
                               & \our & \textbf{49.72} & \textbf{43.51} & \textbf{33.57} & \textbf{40.49} & \colorbox{my_c1!50}{\textbf{41.82} (+2.17\%)} \\
\midrule[0.5pt]
\rowcolor{my_c2!50}
/ & LoRA & 47.57 & 42.45 & 31.76 & 39.28 & 40.27 \\

\midrule[1pt]
&&\multicolumn{4}{c}{\textbf{LLaMA2-13B (INT4)}}& \\ 
\midrule[1pt]
\multirow{2}{*}{\textbf{25\%}} &FedRA  & 47.71 & 53.19 & 35.64 & 43.41 & 44.99 \\ 
& \our & \textbf{53.78} & \textbf{57.26} & \textbf{43.94} & \textbf{47.24} & \colorbox{my_c1!50}{\textbf{50.56} (+5.57\%)} \\

\midrule[0.5pt]

\multirow{2}{*}{\textbf{50\%}} &FedRA  & 50.61 & 54.56 & 37.91 & 44.57 & 46.91 \\ 
& \our & \textbf{55.81} & \textbf{58.09} & \textbf{45.06} & \textbf{47.96} & \colorbox{my_c1!50}{\textbf{51.73} (+4.82\%)} \\

\midrule[0.5pt]

\multirow{2}{*}{\textbf{75\%}} & FedRA  & 52.53 & 55.42 & 39.58 & 45.50 & 48.26 \\ 
& \our & \textbf{57.04} & \textbf{58.67} & \textbf{46.15} & \textbf{48.59} & \colorbox{my_c1!50}{\textbf{52.61} (+4.35\%)} \\

\midrule[0.5pt]
\rowcolor{my_c2!50}
/ & LoRA& 52.40 & 55.45 & 40.33 & 46.14 & 48.58 \\

\bottomrule[1pt]
\end{tabular}
}

%% file: Section/2_Related_Work.tex
\section{Related Work}


\textbf{Layer Pruning.} Layer pruning~\citep{gromov2024unreasonable} has emerged as a prominent technique for LLM compression by eliminating redundant layers. ShortGPT~\citep{men2024shortgpt} reveals that straightforward layer pruning can achieve comparable performance to sophisticated width pruning methods. LaCo~\citep{men2024shortgpt} proposes a concise layer-wise structured pruning method that consolidates posterior layers into their preceding counterparts for model compression. Different from existing works that focus on obtaining a compact model, \our employs layer pruning to address the memory constraints in federated fine-tuning.



\textbf{Memory-Efficient Federated Fine-tuning.} 
Existing memory-efficient federated fine-tuning methods mainly focus on reducing LoRA parameters or intermediate activations.
For instance, FLoRA~\citep{wang2024flora} assigns LoRA ranks based on device resources, while FedKSeed~\citep{qin2023federated} uses zeroth-order optimization to avoid storing activations.
However, these approaches fail to address the memory constraints as they still require retaining the full model in memory. In contrast, \our addresses this bottleneck by strategically pruning layers to reduce model parameters, thereby fundamentally mitigating memory constraints.


%% file: Section/5_Conclusion.tex
\section{Conclusion}

In this paper, we introduce \our, a novel federated fine-tuning paradigm that employs layer pruning to overcome the memory constraints of participating devices. To systematically coordinate the pruning process across devices, we develop a macro-micro synergistic pruning framework. Furthermore, we propose a fine-grained component-level pruning framework to precisely preserve essential components. Extensive experiments on benchmark datasets demonstrate that both \our and its fine-grained variant consistently outperform state-of-the-art baseline methods.

%% file: Table/table_macro_micro_appendix.tex
\begin{table}[!t]
\caption{{Layer grouping results, intra-group layer selection probabilities, and selected layers for submodel construction at round 6.}}
\label{table_macro_micro_round6}
\centering
\resizebox{\linewidth}{!}{
\begin{tabular}{ccccccc}
\toprule[1pt]
{\textbf{\shortstack{Group}}}& {\textbf{The Layer Index within}}  & {\textbf{Layer Selection Probability within}} & \multirow{2}{*}{\textbf{Final Selected Layer}}\\

\textbf{Index} & \textbf{the Group} & \textbf{the Group}  & \\ 

\midrule[1pt]
$\mathcal{G}_{1}$ & [1] & [1.0] & layer 1\\
\midrule
$\mathcal{G}_{2}$ & [2] & [1.0] & layer 2\\
\midrule
$\mathcal{G}_{3}$ & [3] & [1.0] & layer 3\\
\midrule
$\mathcal{G}_{4}$ & [4] & [1.0] & layer 4\\
\midrule
$\mathcal{G}_{5}$ & [5] & [1.0] & layer 5\\
\midrule

\multirow{2}{*}{$\mathcal{G}_{6}$} & [6,7,8,9,10,11,12,13,14,15,16, & [0.05, 0.05, 0.05, 0.05, 0.05, 0.05, 0.05, 0.05, 0.04, 0.04, 0.03, & \multirow{2}{*}{layer 13}\\
& 17,18,19,20,21,22,23,24,25,26] & 0.04, 0.05, 0.05, 0.05, 0.05, 0.05, 0.05, 0.05, 0.05, 0.05] & \\
\midrule

$\mathcal{G}_{7}$ & [27] & [1.0] & layer 27\\
\midrule

$\mathcal{G}_{8}$ & [28] & [1.0] & layer 28\\
\midrule

$\mathcal{G}_{9}$ & [29] & [1.0] & layer 29\\
\midrule

$\mathcal{G}_{10}$ & [30] & [1.0] & layer 30\\
\midrule

$\mathcal{G}_{11}$ & [31] & [1.0] & layer 31\\
\midrule

$\mathcal{G}_{12}$ & [32] & [1.0] & layer 32\\

\bottomrule[1pt]
\end{tabular}
}
\label{table_memory_constraints}
\end{table}

%% file: Table/table_macro_micro_appendix2.tex
\begin{table}[!t]
\caption{{Layer grouping results, intra-group layer selection probabilities, and selected layers for submodel construction at round 194.}}

\label{table_macro_micro_round194}
\centering
\renewcommand{\arraystretch}{1}  
\resizebox{\linewidth}{!}{
\begin{tabular}{ccccccc}
\toprule[1pt]
{\textbf{\shortstack{Group}}}& {\textbf{The Layer Index within}}  & {\textbf{Layer Selection Probability within}} & \multirow{2}{*}{\textbf{Final Selected Layer}}\\

\textbf{Index} & \textbf{the Group} & \textbf{the Group}  & \\ 

\midrule[1pt]
$\mathcal{G}_{1}$ & [1,2,3,4,5] & [0.16, 0.18, 0.21, 0.23, 0.22] & layer 4\\
\midrule
$\mathcal{G}_{2}$ & [6,7] & [0.55, 0.45] & layer 6\\
\midrule
$\mathcal{G}_{3}$ & [8] & [1.0] & layer 8\\
\midrule
$\mathcal{G}_{4}$ & [9] & [1.0] & layer 9\\
\midrule
$\mathcal{G}_{5}$ & [10] & [1.0] & layer 10\\
\midrule

$\mathcal{G}_{6}$ & [11,12,13,14,15,16,17,18,19] & [0.11, 0.11, 0.11, 0.11, 0.12, 0.11, 0.11, 0.11, 0.11] & layer 16\\
\midrule

$\mathcal{G}_{7}$ & [20] & [1.0] & layer 20\\
\midrule

$\mathcal{G}_{8}$ & [21] & [1.0] & layer 21\\
\midrule

$\mathcal{G}_{9}$ & [22] & [1.0] & layer 22\\
\midrule

$\mathcal{G}_{10}$ & [23,24] & [0.48,0.52] & layer 24\\
\midrule

$\mathcal{G}_{11}$ & [25,26] & [0.46,0.54] & layer 25\\
\midrule

$\mathcal{G}_{12}$ & [27,28,29,30,31,32] & [0.16, 0.18, 0.20, 0.17, 0.15, 0.14] & layer 31\\

\bottomrule[1pt]
\end{tabular}
}
\label{table_memory_constraints}
\end{table}

%% file: Table/table_fine_grained_appendix1.tex
\begin{table}[!t]

\caption{Fine-grained layer grouping results, intra-group layer selection probabilities, and selected layers for submodel construction at round 6, where $n^{M}$ and $n^{N}$ denote the MHA and FFN components of the $n$-th layer, respectively. The red highlights indicate novel assembly patterns emerging in submodel construction.}
\label{table_fine_grained_round6}
\centering
\renewcommand{\arraystretch}{0.8} 

\resizebox{\linewidth}{!}{
\begin{tabular}{ccccccc}
\toprule[1pt]
{\textbf{\shortstack{Group}}}& {\textbf{The Layer Index within}}  & {\textbf{Layer Selection Probability within}} & \multirow{2}{*}{\textbf{Final Selected Layer}}\\

\textbf{Index} & \textbf{the Group} & \textbf{the Group}  & \\ 

\midrule[1pt]
$\mathcal{G}_{1}$ & [1$^{M}$] & [1.0] & 1$^{M}$\\
\midrule
$\mathcal{G}_{2}$ & [1$^{N}$] & [1.0] & 1$^{N}$\\
\midrule
$\mathcal{G}_{3}$ & [2$^{M}$] & [1.0] & 2$^{M}$\\
\midrule
$\mathcal{G}_{4}$ & [2$^{N}$] & [1.0] & 2$^{N}$\\
\midrule
$\mathcal{G}_{5}$ & [3$^{M}$] & [1.0] & 3$^{M}$\\
\midrule
$\mathcal{G}_{6}$ & [3$^{N}$] & [1.0] & 3$^{N}$\\
\midrule
$\mathcal{G}_{7}$ & [4$^{M}$] & [1.0] & 4$^{M}$\\
\midrule
$\mathcal{G}_{8}$ & [4$^{N}$] & [1.0] & 4$^{N}$\\
\midrule
$\mathcal{G}_{9}$ & [5$^{M}$] & [1.0] & 5$^{M}$\\
\midrule
$\mathcal{G}_{10}$ & [5$^{N}$] & [1.0] & 5$^{N}$\\
\midrule
$\mathcal{G}_{11}$ & [6$^{M}$] & [1.0] & \textcolor{red}{6$^{M}$}\\
\midrule
\multirow{4}{*}{$\mathcal{G}_{12}$} & [6$^{N}$,7$^{M}$,7$^{N}$,8$^{M}$,8$^{N}$,9$^{M}$,9$^{N}$,10$^{M}$,10$^{N}$,11$^{M}$,11$^{N}$,12$^{M}$,12$^{N}$, & [0.025, 0.025, 0.025, 0.025, 0.025, 0.025, 0.025, 0.025, 0.025, 0.025, 0.025, & \multirow{4}{*}{\textcolor{red}{16$^{N}$}}\\
& 13$^{M}$,13$^{N}$,14$^{M}$,14$^{N}$,15$^{M}$,15$^{N}$,16$^{M}$,16$^{N}$,17$^{M}$,17$^{N}$, & 0.025, 0.025, 0.025, 0.025, 0.025, 0.024, 0.023, 0.023, 0.022, 0.022, \\
& 18$^{M}$,18$^{N}$,19$^{M}$,19$^{N}$,20$^{M}$,20$^{N}$,21$^{M}$,21$^{N}$,22$^{M}$,22$^{N}$, & 0.021, 0.022, 0.022, 0.023, 0.023, 0.025, 0.025, 0.025, 0.025, 0.025, & \\
& 23$^{M}$,23$^{N}$,24$^{M}$,24$^{N}$,25$^{M}$,25$^{N}$,26$^{M}$,26$^{N}$] & 0.025, 0.025, 0.025, 0.025, 0.025, 0.025, 0.025, 0.025, 0.025, 0.025] \\
\midrule

$\mathcal{G}_{13}$ & [27$^{M}$] & [1.0] & 27$^{M}$\\
\midrule

$\mathcal{G}_{14}$ & [27$^{N}$] & [1.0] & 27$^{N}$\\
\midrule

$\mathcal{G}_{15}$ & [28$^{M}$] & [1.0] & 28$^{M}$\\
\midrule

$\mathcal{G}_{16}$ & [28$^{N}$] & [1.0] & 28$^{N}$\\
\midrule

$\mathcal{G}_{17}$ & [29$^{M}$] & [1.0] & 29$^{M}$\\
\midrule

$\mathcal{G}_{18}$ & [29$^{N}$] & [1.0] & 29$^{N}$\\

\midrule

$\mathcal{G}_{19}$ & [30$^{M}$] & [1.0] & 30$^{M}$\\

\midrule

$\mathcal{G}_{20}$ & [30$^{N}$] & [1.0] & 30$^{N}$\\

\midrule

$\mathcal{G}_{21}$ & [31$^{M}$] & [1.0] & 31$^{M}$\\

\midrule

$\mathcal{G}_{22}$ & [31$^{N}$] & [1.0] & 31$^{N}$\\

\midrule

$\mathcal{G}_{23}$ & [32$^{M}$] & [1.0] & 32$^{M}$\\

\midrule

$\mathcal{G}_{24}$ & [32$^{N}$] & [1.0] & 32$^{N}$\\

\bottomrule[1pt]
\end{tabular}
}
\label{table_memory_constraints}
\end{table}

%% file: Table/table_fine_grained_appendix2.tex
\begin{table}[!t]

\caption{Fine-grained layer grouping results, intra-group layer selection probabilities, and selected layers for submodel construction at round 194, where $n^{M}$ and $n^{N}$ denote the MHA and FFN components of the $n$-th layer, respectively. The red highlights indicate novel assembly patterns emerging in submodel construction.}
\label{table_fine_grained_round194}
\centering
\renewcommand{\arraystretch}{0.8} 

\resizebox{\linewidth}{!}{
\begin{tabular}{ccccccc}
\toprule[1pt]
{\textbf{\shortstack{Group}}}& {\textbf{The Layer Index within}}  & {\textbf{Layer Selection Probability within}} & \multirow{2}{*}{\textbf{Final Selected Layer}}\\

\textbf{Index} & \textbf{the Group} & \textbf{the Group}  & \\ 

\midrule[1pt]
$\mathcal{G}_{1}$ & [1$^{M}$,1$^{N}$,2$^{M}$,2$^{N}$,3$^{M}$] & [0.16, 0.18, 0.21, 0.23, 0.22] & \textcolor{red}{2$^{N}$}\\
\midrule
$\mathcal{G}_{2}$ & [3$^{N}$,4$^{M}$,4$^{N}$,5$^{M}$,5$^{N}$] & [0.23,0.22,0.16,0.18,0.21] & \textcolor{red}{4$^{M}$}\\
\midrule
$\mathcal{G}_{3}$ & [6$^{M}$,6$^{N}$] & [0.55, 0.45] & \textcolor{red}{6$^{M}$}\\
\midrule
$\mathcal{G}_{4}$ & [7$^{M}$,7$^{N}$] & [0.52, 0.48] & \textcolor{red}{7$^{M}$}\\
\midrule
$\mathcal{G}_{5}$ & [8$^{M}$] & [1.0] & \textcolor{red}{8$^{M}$}\\
\midrule
$\mathcal{G}_{6}$ & [8$^{N}$] & [1.0] & 8$^{N}$\\
\midrule
$\mathcal{G}_{7}$ & [9$^{M}$,9$^{N}$] & [0.53, 0.47] & \textcolor{red}{9$^{M}$}\\
\midrule
$\mathcal{G}_{8}$ & [10$^{M}$] & [1.0] & \textcolor{red}{10$^{M}$}\\
\midrule
$\mathcal{G}_{9}$ & [10$^{N}$] & [1.0] & \textcolor{red}{10$^{N}$}\\
\midrule
\multirow{2}{*}{$\mathcal{G}_{10}$} & [11$^{M}$,11$^{N}$,12$^{M}$,12$^{N}$,13$^{M}$,13$^{N}$,14$^{M}$,14$^{N}$,15$^{M}$,15$^{N}$, & [0.058, 0.058, 0.058, 0.058, 0.058, 0.059, 0.060, 0.060, 0.062, 0.060,  & \multirow{2}{*}{\textcolor{red}{15$^{M}$}}\\
& 16$^{M}$,16$^{N}$,17$^{M}$,17$^{N}$,18$^{M}$,18$^{N}$,19$^{M}$] & 0.060, 0.059, 0.058, 0.058, 0.058, 0.058, 0.058] &\\
\midrule
$\mathcal{G}_{11}$ & [19$^{N}$] & [1.0] & \textcolor{red}{{19$^{N}$}}\\
\midrule

$\mathcal{G}_{12}$ & [20$^{M}$] & [1.0] & {20$^{M}$}\\

\midrule

$\mathcal{G}_{13}$ & [20$^{N}$] & [1.0] & 20$^{N}$\\
\midrule

$\mathcal{G}_{14}$ & [21$^{M}$] & [1.0] & 21$^{M}$\\
\midrule

$\mathcal{G}_{15}$ & [21$^{N}$] & [1.0] & 21$^{N}$\\
\midrule

$\mathcal{G}_{16}$ & [22$^{M}$] & [1.0] & 22$^{M}$\\
\midrule

$\mathcal{G}_{17}$ & [22$^{N}$] & [1.0] & \textcolor{red}{22$^{N}$}\\
\midrule

$\mathcal{G}_{18}$ & [23$^{M}$,23$^{N}$] & [0.51, 0.49] & \textcolor{red}{23$^{N}$}\\

\midrule

$\mathcal{G}_{19}$ & [24$^{M}$,24$^{N}$] & [0.46, 0.54] & \textcolor{red}{24$^{N}$}\\

\midrule

$\mathcal{G}_{20}$ & [25$^{M}$] & [1.0] & 25$^{M}$\\

\midrule

$\mathcal{G}_{21}$ & [25$^{N}$] & [1.0] & \textcolor{red}{25$^{N}$}\\

\midrule

$\mathcal{G}_{22}$ & [26$^{M}$,26$^{N}$] & [0.43, 0.57] & \textcolor{red}{26$^{N}$}\\

\midrule

$\mathcal{G}_{23}$ & [27$^{M}$,27$^{N}$,28$^{M}$,28$^{N}$,29$^{M}$,29$^{N}$] & [0.15, 0.18, 0.20, 0.17, 0.16, 0.14] & \textcolor{red}{28$^{M}$}\\

\midrule

$\mathcal{G}_{24}$ & [30$^{M}$,30$^{N}$,31$^{M}$,31$^{N}$,32$^{M}$,32$^{N}$] & [0.16, 0.17, 0.20, 0.18, 0.15, 0.14] & \textcolor{red}{31$^{N}$}\\

\bottomrule[1pt]
\end{tabular}
}
\label{table_memory_constraints}
\end{table}